\documentclass[twocolumn,a4]{emulateapj}

\begin{document}

\title{The 3rd IBIS/ISGRI soft gamma-ray survey catalog
\footnote{\rm Based on observations with INTEGRAL, an ESA project
with instruments and science data centre funded by ESA member states
(especially the PI countries: Denmark, France, Germany, Italy,
Switzerland, Spain), Czech Republic and Poland, and with the
participation of Russia and the USA.}}

\author{A. J. Bird\altaffilmark{a}\email{ajb@astro.soton.ac.uk},
  A. Malizia\altaffilmark{b},
  A. Bazzano\altaffilmark{c},
  E. J. Barlow\altaffilmark{a},
  L. Bassani\altaffilmark{b},
  A. B. Hill\altaffilmark{a},
  G. B\'{e}langer\altaffilmark{d},
  F. Capitanio\altaffilmark{a,c},
  D. J. Clark\altaffilmark{a},
  A. J. Dean\altaffilmark{a},
  M. Fiocchi\altaffilmark{c},
  D. G\"{o}tz\altaffilmark{d},
  F. Lebrun\altaffilmark{d},
  M. Molina\altaffilmark{a},
  N. Produit\altaffilmark{f,g},
  M. Renaud\altaffilmark{d},
  V. Sguera\altaffilmark{a},
  J. B. Stephen\altaffilmark{b},
  R. Terrier\altaffilmark{e},
  P. Ubertini\altaffilmark{c},
  R. Walter\altaffilmark{f,g},
  C. Winkler\altaffilmark{h},
  J. Zurita\altaffilmark{d}
}

\altaffiltext{a}{School of Physics and Astronomy, University of Southampton, SO17 1BJ, UK} 
\altaffiltext{b}{IASF-INAF, Via Gobetti 101, 40129 Bologna, Italy} 
\altaffiltext{c}{IASF-INAF, Via Fosso del Cavaliere 100, 00133 Rome, Italy}
\altaffiltext{d}{CEA-Saclay, DAPNIA/Service d'Astrophysique, F91191, Gif sur Yvette Cedex, France}
\altaffiltext{e}{Federation de recherche APC, College de France 11, place Marcelin Berthelot, F75231, Paris, France}
\altaffiltext{f}{Geneva Observatory, University of Geneva, Chemin des Maillettes 51, CH--1290 Sauverny, Switzerland}
\altaffiltext{g}{INTEGRAL Science Data Centre, Chemin d'Ecogia 16, CH--1290 Versoix, Switzerland}
\altaffiltext{h}{ESA-ESTEC,Research and Scientific Support Dept., Keplerlaan 1, 2201 AZ, Noordwijk, The Netherlands}


\begin{abstract}

  In this paper we report on the third soft gamma-ray source catalog
  obtained with the IBIS/ISGRI gamma-ray imager on board the INTEGRAL
  satellite. The scientific dataset is based on more than 40 Ms of
  high quality observations performed during the first three and a
  half years of Core Program and public IBIS/ISGRI
  observations. Compared to previous IBIS/ISGRI surveys, this catalog
  includes a substantially increased coverage of extragalactic
  fields, and comprises more than 400 high-energy sources
  detected in the energy range 17--100 keV, including both transients
  and faint persistent objects which can only be revealed with longer
  exposure times.

\end{abstract}



\section{Introduction}

Since its launch in 2002, the INTEGRAL (International Gamma-Ray
Astrophysics Laboratory) observatory has carried out more than 4 years
of observations in the energy range from 5 keV -- 10 MeV. INTEGRAL is
an observatory-type mission, and most of the total observing time
(65\% in the nominal phase, 75\% during the mission extension) is
awarded as the General Programme to the scientific community at
large. Typical observations last from 100 ks up to two weeks. As a
return to the international scientific collaborations and individual
scientists who contributed to the development, design and procurement
of INTEGRAL, a part of the observing time (from 35\% to 25\%) is
allocated to the Core Programme. During the the nominal lifetime, this
programme consisted of three elements, a deep exposure of the Galactic
central radian, regular scans of the Galactic Plane, pointed
observations of the Vela region and Target of Opportunity follow-ups
\citep{Winkler2003}.

The IBIS (Imager on Board INTEGRAL spacecraft) imaging instrument is
optimised for survey work with a large ($30^{\circ}$) field of view
with excellent imaging and spectroscopy capability
\citep{Ubertini2003}, and has formed the basis of several previous
INTEGRAL surveys.

The frequent Galactic Plane Scans (GPS) within the Core Programme,
performed in the first year of operations, were successfully exploited
to yield a first survey of the galactic plane to a depth of $\sim$1
mCrab in the central radian \citep{Bird2004cat1}. This gave evidence
of a soft gamma-ray sky populated with more than 120 sources,
including a substantial fraction of previously unseen sources. The
second IBIS/ISGRI catalog \citep{Bird2006cat2} used a greatly
increased dataset (of $\sim$10Ms) to unveil a soft gamma-ray sky
comprising 209 sources, again with a substantial component
($\sim$25\%) of new and unidentified sources.

\section{The IBIS `all sky' survey}

In this paper we provide the third IBIS/ISGRI soft gamma-ray survey
catalog, comprising more than 400 high-energy sources. 

The instrumental details and sensitivity can be found in
\citet{Lebrun2003} and \citet{Ubertini2003}. The data are collected
with the low-energy array, ISGRI (INTEGRAL Soft Gamma-Ray Imager;
\cite{Lebrun2003}), consisting of a pixellated 128x128 CdTe
solid-state detector that views the sky through a coded aperture
mask. IBIS/ISGRI generates images of the sky with a 12$'$ (FWHM)
resolution and arcmin source location accuracy over a $\sim
19^{\circ}$(FWHM) field of view in the energy range 15--1000 keV.

This `all sky' catalog uses mosaic image data from the first 3.5 years
of IBIS/ISGRI Core Programme and public observations. The dataset used
in this catalog ensures that $>$70\% of the sky is now observed with
an exposure of at least 10ks (see Figure~\ref{fig:expo}). As for
previous catalogs, the aim is to provide a prompt release of
information to the community.


\section{Data analysis and catalog construction}

The methods used for production of this catalog are the same as, or
close derivatives of, those used in the second IBIS/ISGRI catalog
production \citep{Bird2006cat2}. Refinements have been made in various
areas, and some techniques have been extended to deal with the larger
dataset now in use.

INTEGRAL/IBIS data is organised in short pointings referred to as
science windows (scw) each of typically 2000s. During the majority of
observations, the INTEGRAL telescope axis is dithered around the
nominal pointing direction by a few degrees in order to aid image
reconstruction. In this observing mode, science windows contain data
taken either during the pointings or during the short slews between
the dither positions; pointing and slew data is not mixed within a
science window. A small fraction of the observations are performed in
`staring' mode, where the telescope axis is kept fixed on a target for
long periods without dithering. In this case, the long pointing is
divided into several science windows, but no slews are present. 

\subsection{Input dataset and pipeline processing \label{pipeline}}

The survey input dataset consists of all pointing data available at
the end of May 2006, from revolutions (orbits) 12-429 inclusive,
covering the time period from launch to the end of April 2006. This
results in more than 40~Ms exposure time in this iteration of the
survey analysis.

Pipeline processing was carried out using the latest version of the
standard INTEGRAL analysis software (OSA 5.1; \cite{Goldwurm2003}) up to
the production of sky images for individual science windows. Five
primary search bands (17--30, 18--60, 20--40, 30--60 and 40--100 keV)
were used to both optimise the source search sensitivity and provide
compatibility with previous datasets. The new version of the OSA
software allowed lower energy thresholds to be used than in previous
catalogs, improving the sensitivity to sources with very soft spectra.

A catalog of known or expected sources is a key input for the image
deconvolution process. The final input catalog used in the image
processing described here comprised $\sim$350 excesses produced
primarily by a preliminary processing of a smaller dataset using {\em
  OSA 5.1} and an input catalog based on the second IBIS/ISGRI survey,
plus all other INTEGRAL-detected sources (i.e. those with an IGR
designation) published up to the start of processing.

\subsection{Science window selection}

When constructing a final mosaic of all images, it is important to
remove the small fraction of images for which the image deconvolution
process has not been successful. These mainly include data taken
during or following severe solar activity or near spacecraft perigee
passages when the background modelling is difficult.

As for the second IBIS/ISGRI catalog production, the image rms (after
removal of sources) was used as the primary indicator of image
quality. The image rms was determined for each significance map (at
science window level), and the distribution of the image rms
statistics for all science windows was determined. The mean and
variance of this distribution was determined in order to define what
can be considered a `good' image rms. An acceptance threshold was then
set at 2 sigma above the mean image rms, and any individual images with higher
rms than this were discarded. Typically, this resulted in any
image with an rms greater than 1.08 (after removal of sources) being
rejected.  Of the $\sim$24000 scw processed, $\sim$20000 scw were
retained in the final scw list.

Additionally, science windows acquired in `staring' mode, and data
taken during the instrument Performance Verification (PV) phase (for
simplicity, this was taken as up to and including the calibration
activities in revolution 45) were removed from the main science window
lists due to their potential adverse effect on final mosaic
quality. Separate science window lists for staring data and PV data
were constructed with higher rms limits to allow for the poorer image
quality.

\subsection{Mosaic construction}

The selected science windows were mosaiced using a proprietary tool
optimised to create all-sky galactic maps based on several thousand
input science windows.

The higher exposure and long timebase spanned by this latest dataset
has introduced a new problem since the second catalog. The source
search methods we employ are optimised for detection of persistent
flux from a source; a highly variable source may be clearly detectable
during outburst, while having an undetectably low mean flux over the
full dataset. In order to compensate for this problem, for this third
catalog, we constructed mosaics over three timescales. Maps were
created for each {\em revolution} which contained valid data. This is
optimised to detect sources active on timescales of the order of a
day.  We identified 26 sequences of consecutive revolutions which had
similar pointings. Thus these {\em revolution sequences} could best be
analysed as a single observation, and sensitivity for sources on
longer timescales than revolutions (i.e. order of weeks) could be
optimised. Ultimately, persistent sources can best be detected in an
{\em all-archive} accumulation of all available high-quality data.

Maps were created for each of these timescales, in each of the five
energy bands described in section~\ref{pipeline}, these being chosen to
provide both coverage of the most sensitive energy range for ISGRI and
sensitivity to various typical source emission profiles.

For each energy band and time period all-sky mosaics were made in four
projections: centred on galactic centre, centred on galactic
anti-centre, north galactic polar and south galactic polar. The
purpose of these multiple projections is to present the automatic
source detection algorithms with source PSFs with the minimum
possible distortions.

\begin{figure}
\plotone{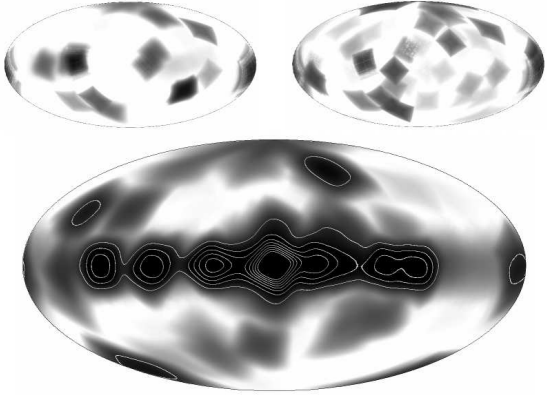}
\caption{Exposure maps for the third IBIS/ISGRI catalog observations:
  (lower) the all-archive mosaic (contour levels are at 500 ksec)
  which excludes (top left) performance verification phase exposure
  and (top right) staring mode data \label{fig:expo}}
\end{figure}

Additionally, maps were made of the two special datasets: the {\em
  staring mode data}, which was excluded from the main mosaics due to
the presence of stronger image artefacts which impede the source
search algorithms; the {\em performance verification (PV) data}, which
has poorer image quality and would have been largely rejected by the
standard rms filters, but contains significant exposure on parts of
the sky otherwise poorly exposed. The exposure maps for these three
separate datasets are shown in Figure~\ref{fig:expo}.

In total $\sim$750 maps were created and searched.  Each mosaic used a
pixel size of 0.04$^{\circ}$ (2.4$'$) at the centre of the mosaic in
order to optimise source detection and location over the whole mosaic.

\subsection{Source searching and location}

Each of the mosaics was searched using the {\it SExtractor 2.4.4}
software \citep{sextractor}. The source
positions measured by {\it SExtractor} represent the centroid of the
source calculated by taking the first order moments of the source
profile (referred to by {\it SExtractor} as the barycenter method).

Source detectability is limited at the faintest levels by background
noise and can be improved by the application of a linear filtering of
the data.  In addition, source confusion in crowded fields can be
minimised by the application of a bandpass filter, specifically the
{\it mexhat} bandpass filter is used in the {\it SExtractor} software.
The convolution of the filter with the mosaic alters the source
significances, hence {\it SExtractor} uses the source positions
identified from the filtered mosaic to extract the source
significances from the original mosaic.

Additional manual checks were performed on each map to check for the
(rare) occasions where {\it SExtractor} fails due to the close
proximity of two sources.

\subsection{Source list filtering}

An initial list of 815 excesses was generated by integration of all
lists derived from mosaic images on whole-archive, revolution sequence
and revolution timescales.

\begin{figure}[htbp]
\centering
\plotone{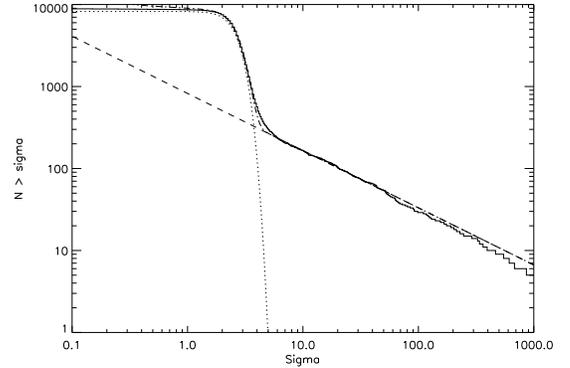}
\vspace{0.5cm}
\caption{Distribution of significances of source-like excesses found
in the 30-60 keV all-sky mosaic. The solid line represents the data;
the other lines indicate the fit model components, see text for details.
\label{fig:cut}}
\end{figure}

In order to identify an excess as a source it is necessary first to
identify the significance level at which the source population
dominates over the noise distribution.  To this end we produce a
log-log plot of the number of excesses detected by {\it SExtractor} above a
specific significance as a function of that significance.  This is
shown in Figure~\ref{fig:cut} for the 30-60 keV all-sky mosaic.

This distribution is fitted by an integrated Gaussian and power law
function; the power law component represents the underlying source
population and the integrated Gaussian fits to the noise component of
the distribution. This model and its components are illustrated in
Figure~\ref{fig:cut}: the dashed line represents the power-law component;
the dotted line represents the integrated Gaussian component;
the dotted-dashed line represents the overall model.  Based upon the
parameters of the fitted model it is possible to calculate the
significance at which the noise distribution contributes 1\% of the
detected excesses - in the case of the 30-60 keV band mosaic this level is 
4.8$\sigma$. 

However, this cut-off is based upon the global properties of an
individual mosaic and the maps contains systematic errors that are not
uniformly spatially distributed.  The majority of the systematic noise
is attributable to the very brightest sources and crowded regions
where the deconvolution software has problems cleaning all of the
image artefacts.  Consequently the systematic noise is localised to
specific regions of the map. In areas with large amounts of systematic
noise, such as the Galactic Centre, the cut-off significance will be
higher than the global value, while those areas where there is no
noticeable systematic noise, principally the extragalactic sky, a
lower threshold is appropriate.

The situation is further complicated because each map, with a
different energy band, exposure and instrument configuration, will
have a subtly different statistical and systematic noise distribution,
and hence source detection criteria. 

We have applied an initial absolute threshold of 4.5$\sigma$ in the
maps, and all excesses above this threshold were then combined into a
preliminary source list. Thereafter, each candidate source was
visually inspected and checked for appropriate PSF shape, and removal
of systematic map artefacts. When necessary, proprietary tools were
used to compare the peaks found in the maps against local rms levels,
and perform 2-dimensional gaussian fitting to the source PSFs. Both
techniques have been found to aid source extraction in regions of
non-statistical background.

After all selection processes, we obtain a source list containing 421
sources, as shown in Table \ref{table:sourcelist} which have been located in the short and long timescale maps.

We can estimate the number of possible false detections in our source
list as follows. The analysis for the all-archive map shown in
Figure~\ref{fig:cut} indicates that above 5 sigma significance, there
remains a 1\% probability of a false detection, while for the
revolution and revolution sequence maps, the corresponding figure is 6
sigma. In total, 372 of our sources meet one or both of these criteria
and hence we estimate 4 false detections from this sample. The
remaining 49 sources should be treated with more caution, and we
estimate that 10--20\% of these sources may be false detections,
noting in passing that our source inspection processes have already
discarded 75\% of the excesses found between 4.5 and 5 sigma. Hence,
overall, we expect that $<3$\% of the catalog sources result from
false detections, and the majority of these will be sources below 5
sigma (or 6 sigma in the revolution maps).

A number of deep studies have been performed on the $\sim 4^{\circ}
\times 1^{\circ}$ region surrounding the Galactic Centre, which is a
highly variable sky region containing a group of sources which cannot
be de-blended by the imaging capabilities of IBIS alone. For example,
\citet{Belanger2006} provides a study of the Galactic Centre region
using both spectral and temporal analyses which are beyond the scope
of this broader survey.


\section{Notes on the table}

\subsection{Source Positions and Uncertainties\label{positions}}

The astrometric coordinates of the source positions were extracted from
the mosaics by the centroiding routines built into {\it SExtractor
  2.4.4}. The position of each source was taken from the mosaic that
had the most significant detection.  More than 300 of the sources in
the 3$^{rd}$ IBIS/ISGRI catalog have well defined positions in the
SIMBAD/NED database.  Measuring the angular distance between the
measured positions and those provided by the SIMBAD database gives an
indication of the source position errors.

\begin{figure}[htbp]
\centering
\plotone{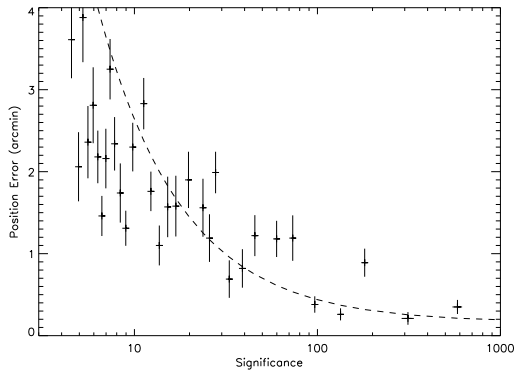}
\caption{The binned mean source position error of sources as a
function of source significance.  Each bin contains 10 sources.  The
dashed line represents the model shown in Equation~\ref{eq:pos}.
\label{fig:pos_err}}
\end{figure}

The point source location error of IBIS is highly dependent upon the
significance of the source detected \citep{Gros2003}. By binning
together sources of similar significance and calculating the mean
source position error we can see how the source position accuracy
varies with significance; this is shown in Figure~\ref{fig:pos_err},
which also shows the empirical model derived by \cite{Gros2003}:

\begin{equation}
  \delta x = 22.1\sigma^{-0.95} + 0.16 
\label{eq:pos}
\end{equation}

$\delta x$ is the error in the source position (90\% confidence, in
arc minutes) and $\sigma$ is the source significance. It is clear that
the positions in this catalog are consistent with the expected
scientific imaging performance of IBIS/ISGRI, and no additional
systematic effects are introduced during our processing.

Compared to the second IBIS/ISGRI catalog, the position
determination has been improved in several ways: the source positions
are {\em always} taken from the mosaic in which they have the greatest
significance; the use of multiple map projections has minimised the
distortion of source PSFs; and the mosaics themselves are
generated at higher resolution.

\subsection{Fluxes and significances}

The fluxes quoted in the table are the time-averaged fluxes over the
whole dataset in two energy bands (20--40 and 40--100 keV). The
significances quoted are the highest significance in any single map
(also identified in the table), since this gives the best indication
of the robustness of source detection. However, it should be noted
therefore that the flux and significance values may derive from very
different subsets of the data, and may initially appear contradictory.


\section{Discussion}

We have derived an `unbiased' catalog of 421 sources observed in a
systematic analysis of the IBIS/ISGRI Core Programme and public data
spanning nearly 3.5 years of operations.

\begin{figure}[htbp]
\includegraphics[angle=-90,scale=0.3]{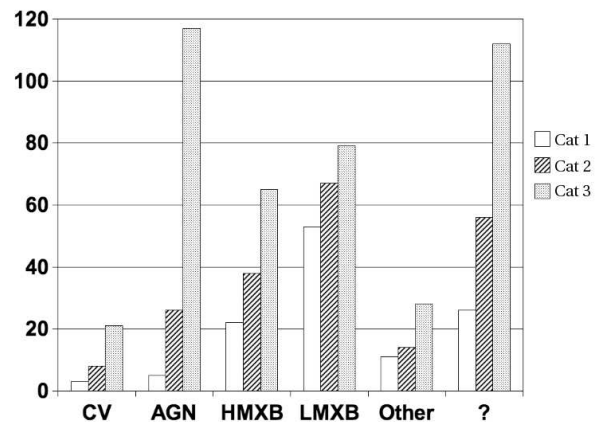}
\caption{Numbers of sources in the 1st, 2nd and 3rd IBIS/ISGRI catalogs, 
classified by type.
\label{fig:compare1}}
\end{figure}

Figure~\ref{fig:compare1} illustrates a simple breakdown of the
sources presented in this catalog by source type, and how this
breakdown compares to previous catalogs. Note that we only include
firm source type determinations in the following analysis, sources are
not characterised based on their hard X-ray characteristics alone, but
based on multi-waveband analysis. In the case of candidate AGN, an
extragalactic nature is strongly indicated by multiwaveband analysis
using radio, infra-red and X-ray archival data, by which their
optical counterpart has been found to be associated with a galaxy.

This catalog is composed of 421 sources of which 171 are galactic
accreting systems (corresponding to 41\%), 122 are extragalactic
objects (29\%), 15 are of other types, and 113 (26\%) are still
not firmly classified.  

Compared to the second catalog \citep{Bird2006cat2}, the most dramatic
change is the increase in AGN number, largely due to the increased
exposure away from the Galactic Plane. This is also reflected in the
increased CV detections, although in this case, it is because we are
sampling a local approximately spherical distribution of objects
within $\sim$400pc.

For galactic sources, this implies we have now detected more than half
of the sources reported in the 2-10 keV band in the catalogs of
\citet{Liu2000} and \citet{Liu2001}. Taking the HMXB as an example,
the number of known HMXB was 30 in 1983 \citep{Paradijs1983},
increasing to 69 in 1995 \citep{Paradijs1995}, and increasing
further to 130 in 2001 \citep{Liu2001}. Most of the new sources were
identified with Be/X ray binaries and some were only tentatively
identified as High or Low Mass on the basis of transient
characteristic or spectral behaviour. This catalog, extending up to
100 keV, includes 68 firm HMXBs, implying not only detection of known
sources but also that a large number (19 in total) of the new INTEGRAL
sources are being identified with such systems.  This is somewhat
different to the results of 6 years of BeppoSAX/WFC operations, which
detected predominantly outbursts from LMXB systems. While INTEGRAL
continues to detect LMXBs, the rate of discovery is much lower than for
the high-mass systems.

The percentage of sources without a firm identification has remained
almost constant since the first IBIS/ISGRI catalog, at
$\sim$25\%. This is despite an active and successful campaign of
follow-up observations in other wavebands (\citet{Masetti4} and
references therein). Of the 110 unclassified sources, around 25\% have
unconfirmed or `tentative' classifications.  INTEGRAL Gamma-ray (IGR)
sources, represent detections that are either entirely new or those
with no obvious counterpart or association in the hard X-ray and/or
gamma-ray wavebands. There is a total of 167 IGRs in the third
catalog, of which 69 have been firmly classified, predominantly as
AGN, HMXB and intermediate polar CVs.  The percentage of IGRs which
have now been classified rises to $\sim$55\% if the tentative
classifications are included.

Finally, we point out some interesting features of this catalog.
Firstly, we have detected 21 CVs of which 9 are new detections and
remarkably, for most of them emission is extended up to 100
keV. Secondly, we see the emergence of the supergiant fast X-ray
transient (SFXT) class. The search for identification of IGRs sources
has resulted in 8 firm associations and 4 possible ones. Also, there
are at least 3 sources that have been observed not in coincidence with
any recurrence time and reported either by Swift/XRT or Chandra. This
implies the sources are rather persistent HMXB with luminous flares
such as the HMXB supergiant neutron star systems 4U 1907+097 (Atel
915).

Finally, we note that 5 of the sources listed in the second IBIS/ISGRI
catalog are not detected in this analysis, although this may be due to
source variability in at least some cases.

\subsection{Concluding comments}

The positions derived from IBIS are forming the basis for an active
program of follow-up observations in other wavebands, mainly X-ray
(XMM-Newton, Chandra, RXTE and Swift), optical, IR and radio. The IBIS
survey team, including scientists from five different institutes, will
continue to refine the analysis techniques and apply them to the
ever-increasing IBIS dataset. Further catalogs are expected to be
released whenever the dataset and/or analysis tools justify them.

This is a golden age for high energy astronomy. The survey
capabilities of IBIS/ISGRI and Swift/BAT are providing exceptional
coverage of the soft gamma-ray sky, and intriguing links are now being
found with the TeV sky being explored by CANGAROO, HESS, MAGIC and
VERITAS. These telescopes will soon be joined by AGILE and GLAST,
providing coverage over 15 orders of magnitude in energy from keV to
TeV. Furthermore, the sources now being discovered will form the vital
input catalogs for the next generations of narrow-field instruments
such as Simbol-X in the X-ray domain and GRI at gamma-ray energies.


\acknowledgments

We acknowledge the following funding: in Italian Space Agency
financial and programmatic support via contracts I/R/046/04 and
ASI/INAF I/023/05/0; in UK via PPARC grant PP/C000714/1; in France,
we thank CNES for support during ISGRI development and INTEGRAL data
analysis.  This research has made use of: data obtained from the High
Energy Astrophysics Science Archive Research Center (HEASARC) provided
by NASA's Goddard Space Flight Center; the SIMBAD database operated at
CDS, Strasbourg, France; the NASA/IPAC Extragalactic Database (NED)
operated by the Jet Propulsion Laboratory, California Institute of
Technology, under contract with the National Aeronautics and Space
Administration.


\clearpage
\LongTables

\begin{deluxetable*}{p{4cm}ccccclccl} 
\tabletypesize{\scriptsize} 
\tablecaption{3rd IBIS/ISGRI Catalog \label{table:sourcelist}} 
\tablewidth{0pt} 
\tablehead{ 
\colhead{Name\tablenotemark{a}} & 
\colhead{RA} & 
\colhead{Dec} & 
\colhead{Error\tablenotemark{b}} & 
\colhead{F20-40\tablenotemark{c}} & 
\colhead{F40-100\tablenotemark{c}} & 
\colhead{Type\tablenotemark{d}} & 
\colhead{Signif\tablenotemark{e}} & 
\colhead{Exposure\tablenotemark{f}} & 
\colhead{MapCode\tablenotemark{g}} 
} 
\startdata 
{\bf IGR J00040+7020} &  1.006 & 70.336 & 3.8 &  0.8$\pm$0.1 &  0.9$\pm$0.3 & AGN? &   6.6 & 1237.0& B5 \\ 
{\bf IGR J00234+6141} &  5.726 & 61.706 & 4.8 &  0.5$\pm$0.1 & $<$ 0.4 & CV,IP &   5.2 & 1841.0& S142B1 \\ 
{\bf IGR J00245+6251} &  6.115 & 62.843 & 2.3 &  0.2$\pm$0.1 & $<$ 0.4 & GRB &  11.5 & 1823.0& R266B1 \\ 
{\bf 4U 0022+63} &  6.319 & 64.159 & 3.6 &  0.7$\pm$0.1 &  0.7$\pm$0.2 & SNR &   7.1 & 1756.0& B5 \\ 
{\bf IGR J00256+6821} &  6.394 & 68.348 & 4.5 &  0.5$\pm$0.1 &  1.1$\pm$0.2 & AGN? &   5.5 & 1474.0& B3 \\ 
V709 Cas        &  7.204 & 59.306 & 0.9 &  4.4$\pm$0.1 &  2.5$\pm$0.2 & CV,IP &  36.4 & 1765.0& B5 \\ 
{\bf IGR J00291+5934} &  7.253 & 59.566 & 0.5 &  4.6$\pm$0.1 &  5.1$\pm$0.2 & LMXB,msecXP,T &  78.0 & 1752.0& R262B1 \\ 
{\bf IGR J00333+6122} &  8.360 & 61.457 & 4.0 &  0.7$\pm$0.1 &  0.7$\pm$0.2 & ? &   6.2 & 1777.0& B5 \\ 
{\bf 1ES 0033+59.5} &  8.985 & 59.829 & 2.8 &  1.1$\pm$0.1 &  1.1$\pm$0.2 & AGN,BL Lac &   9.4 & 1729.0& B4 \\ 
{\bf IGR J00370+6122} &  9.264 & 61.371 & 4.1 &  0.5$\pm$0.1 & $<$ 0.4 & HMXB,SG? &   6.1 & 1731.0& R147B1 \\ 
RX J0053.8-7226        & 13.543 & -72.429 & 3.4 &  3.1$\pm$0.4 &  1.6$\pm$0.7 & HMXB,XP,Be,T &   7.7 & 132.0& B1 \\ 
gam Cas        & 14.158 & 60.714 & 0.9 &  4.4$\pm$0.1 &  1.3$\pm$0.2 & HMXB,Be &  34.4 & 1503.0& B4 \\ 
SMC X-1        & 19.283 & -73.448 & 0.4 & 38.4$\pm$0.4 &  8.3$\pm$0.7 & HMXB,XP & 100.2 & 139.0& B4 \\ 
1A 0114+650        & 19.500 & 65.289 & 0.6 &  9.9$\pm$0.1 &  4.8$\pm$0.3 & HMXB,XP &  64.4 & 1187.0& B5 \\ 
{\bf 4U 0115+634} & 19.619 & 63.743 & 0.2 & 42.5$\pm$0.1 & 11.8$\pm$0.3 & HMXB,XP,T & 792.2 & 1189.0& R238B2 \\ 
{\bf IGR J01363+6610} & 24.019 & 66.166 & 3.7 & $<$ 0.4 & $<$ 0.6 & HMXB,Be,T &   6.9 & 916.0& R185B2 \\ 
{\bf 4U 0142+614} & 26.631 & 61.747 & 1.9 &  1.9$\pm$0.2 &  3.9$\pm$0.3 & AXP &  14.7 & 749.0& B3 \\ 
{\bf RX J0146.9+6121} & 26.745 & 61.354 & 2.0 &  2.7$\pm$0.2 &  1.3$\pm$0.3 & HMXB,XP,Be,T? &  13.6 & 732.0& B5 \\ 
{\bf IGR J01528-0326} & 28.254 & -3.441 & 4.0 &  0.9$\pm$0.2 &  2.2$\pm$0.4 & AGN,Sy2 &   6.3 & 575.0& B3 \\ 
{\bf NGC 788} & 30.267 & -6.822 & 1.8 &  2.9$\pm$0.2 &  3.0$\pm$0.4 & AGN,Sy2 &  15.5 & 594.0& B5 \\ 
{\bf IGR J02097+5222} & 32.408 & 52.412 & 5.1 &  1.9$\pm$0.4 &  1.5$\pm$0.7 & AGN,Sy1 &   4.9 & 209.0& B5 \\ 
{\bf SWIFT J0216.3+5128} & 34.137 & 51.431 & 5.0 &  1.3$\pm$0.5 &  2.9$\pm$0.7 & AGN,Sy2 &   5.0 & 193.0& B3 \\ 
{\bf NGC 985} & 38.677 & -8.804 & 4.7 &  0.6$\pm$0.2 &  1.8$\pm$0.4 & AGN,Sy1 &   5.3 & 435.0& B2 \\ 
{\bf GT 0236+610} & 40.145 & 61.242 & 3.6 &  1.6$\pm$0.3 &  2.4$\pm$0.5 & HMXB,microQSO &   7.1 & 324.0& B5 \\ 
{\bf NGC 1052} & 40.241 & -8.242 & 4.4 &  1.3$\pm$0.2 & $<$ 0.9 & AGN,Sy2 &   5.6 & 407.0& B5 \\ 
{\bf RBS 345} & 40.567 &  5.530 & 5.5 &  0.8$\pm$0.3 &  1.5$\pm$0.5 & AGN,Sy1 &   4.5 & 444.0& B5 \\ 
{\bf NGC 1068} & 40.689 &  0.016 & 2.8 &  1.6$\pm$0.2 &  1.9$\pm$0.4 & AGN,Sy2 &   9.2 & 589.0& B3 \\ 
QSO B0241+62        & 41.285 & 62.480 & 2.3 &  3.4$\pm$0.3 &  4.0$\pm$0.6 & AGN,Sy1 &  11.4 & 283.0& B3 \\ 
{\bf IGR J02504+5443} & 42.604 & 54.721 & 4.1 &  1.6$\pm$0.3 &  2.1$\pm$0.5 & AGN? &   6.1 & 285.0& B5 \\ 
{\bf MCG-02-08-014\tablenotemark{h}} & 43.120 & -8.485 & 4.5 &  1.2$\pm$0.3 & $<$ 1.1 & AGN,Sy2 &   5.5 & 330.0& B5 \\ 
{\bf NGC 1142} & 43.771 & -0.204 & 2.5 &  2.7$\pm$0.3 &  3.4$\pm$0.5 & AGN,Sy2 &  10.6 & 402.0& B3 \\ 
{\bf B3 B0309+411B} & 48.273 & 41.343 & 4.1 &  1.9$\pm$0.3 & $<$ 1.2 & AGN,Sy1/RG &   6.1 & 218.0& B5 \\ 
{\bf IGR J03184-0014\tablenotemark{i}} & 49.600 & -0.229 & 4.0 &  3.8$\pm$0.7 & $<$ 2.2 & AGN,QSO/BAL &   6.3 &  81.0& B4 \\ 
{\bf NGC 1275} & 49.953 & 41.517 & 2.6 &  3.1$\pm$0.4 &  1.5$\pm$0.6 & AGN,Sy2 &  10.1 & 211.0& B4 \\ 
{\bf 1H 0323+342} & 51.088 & 34.178 & 4.3 &  2.5$\pm$0.5 &  2.4$\pm$0.9 & AGN,Sy1 &   5.9 &  99.0& B5 \\ 
{\bf GK Per} & 52.778 & 43.934 & 4.1 &  1.4$\pm$0.3 &  1.1$\pm$0.6 & CV,IP &   6.2 & 270.0& R273B2 \\ 
{\bf EXO 0331+530} & 53.741 & 53.169 & 0.2 & 214.6$\pm$0.3 & 44.0$\pm$0.5 & HMXB,XP,Be,T & 920.5 & 333.0& R273B2 \\ 
{\bf IGR J03532-6829} & 58.308 & -68.483 & 3.6 & $<$ 0.9 & $<$ 1.5 & AGN,RG &   7.1 & 151.0& S012B3 \\ 
X Per        & 58.835 & 31.049 & 1.0 & 26.4$\pm$1.0 & 31.8$\pm$1.7 & HMXB,XP,Be &  30.5 &  51.0& B5 \\ 
{\bf 3C 111} & 64.573 & 38.014 & 3.1 &  4.9$\pm$0.7 &  6.1$\pm$1.2 & AGN,Sy1/BLRG &   8.5 &  72.0& B5 \\ 
{\bf LEDA 168563} & 73.028 & 49.514 & 3.0 &  3.6$\pm$0.5 &  3.8$\pm$0.8 & AGN,Sy1 &   8.6 & 135.0& B5 \\ 
{\bf ESO 33-2} & 73.918 & -75.602 & 5.4 &  1.7$\pm$0.3 &  1.3$\pm$0.6 & AGN,Sy2 &   4.5 & 230.0& B3 \\ 
{\bf IGR J05053-7343} & 76.329 & -73.716 & 3.6 &  1.0$\pm$0.3 & $<$ 1.2 & ? &   7.0 & 229.0& R028B1 \\ 
{\bf 4U 0517+17} & 77.676 & 16.477 & 3.0 &  3.6$\pm$0.4 &  3.8$\pm$0.6 & AGN,Sy1.5 &   8.8 & 233.0& B3 \\ 
{\bf IGR J05270-6631} & 81.756 & -66.511 & 4.4 &  1.3$\pm$0.3 & $<$ 1.2 & ? &   5.7 & 207.0& B4 \\ 
LMC X-4        & 83.212 & -66.368 & 0.4 & 48.0$\pm$0.3 & 15.6$\pm$0.6 & HMXB,XP & 144.6 & 203.0& B1 \\ 
Crab        & 83.629 & 22.018 & 0.2 & 1000 & 1000 & PWN,PSR & 4529.5 & 572.0& B5 \\ 
{\bf 1A 0535+262} & 84.732 & 26.358 & 2.8 &  3.0$\pm$0.3 &  2.0$\pm$0.4 & HMXB,XP,Be,T &   9.4 & 404.0& B1 \\ 
{\bf LMC X-1} & 84.903 & -69.749 & 1.1 & $<$ 0.6 & $<$ 1.1 & HMXB,BH &  26.3 & 219.0& S012B3 \\ 
{\bf PSR B0540-69.3} & 85.057 & -69.297 & 3.9 &  1.9$\pm$0.3 &  1.9$\pm$0.6 & XB,XP &   6.5 & 221.0& B3 \\ 
{\bf BY Cam} & 85.737 & 60.850 & 4.8 &  2.7$\pm$0.5 &  2.2$\pm$0.9 & CV,P &   5.2 & 161.0& B4 \\ 
MCG+08-11-011        & 88.717 & 46.442 & 3.8 &  4.4$\pm$0.8 &  3.1$\pm$1.4 & AGN,Sy1.5 &   6.6 &  44.0& B5 \\ 
{\bf SWIFT J0601.9-8636} & 91.838 & -86.571 & 5.2 &  1.4$\pm$0.4 &  2.3$\pm$0.7 & AGN,Sy2? &   4.8 & 152.0& R099B2 \\ 
{\bf PKS 0611-663} & 92.938 & -66.433 & 4.6 &  1.1$\pm$0.3 &  2.2$\pm$0.6 & AGN &   5.5 & 191.0& B3 \\ 
{\bf Mrk 3} & 93.891 & 71.043 & 1.2 &  4.9$\pm$0.2 &  6.8$\pm$0.4 & AGN,Sy2 &  25.1 & 547.0& B5 \\ 
4U 0614+091        & 94.280 &  9.139 & 0.8 & 26.8$\pm$0.7 & 22.1$\pm$1.1 & LMXB,B,A &  39.2 &  83.0& B5 \\ 
{\bf IGR J06239-6052\tablenotemark{j}} & 95.978 & -60.898 & 4.9 &  2.5$\pm$0.5 &  2.1$\pm$0.8 & ? &   5.1 & 112.0& B5 \\ 
{\bf IGR J06253+7334} & 96.340 & 73.602 & 5.2 &  0.8$\pm$0.2 & $<$ 0.9 & CV,IP &   4.7 & 545.0& B4 \\ 
{\bf IGR J06292+4858} & 97.301 & 48.974 & 5.0 &  8.8$\pm$1.8 & $<$ 6.0 & ? &   4.9 &  14.0& B1 \\ 
{\bf Mrk 6} & 103.032 & 74.423 & 2.0 &  2.4$\pm$0.2 &  3.4$\pm$0.4 & AGN,Sy1.5 &  14.0 & 578.0& B5 \\ 
{\bf IGR J07295-1329} & 112.376 & -13.158 & 5.4 &  1.9$\pm$0.5 & $<$ 1.8 & ? &   4.5 & 116.0& B5 \\ 
{\bf IGR J07437-5137} & 115.920 & -51.617 & 5.0 &  1.2$\pm$0.2 & $<$ 0.7 & ? &   5.0 & 548.0& B5 \\ 
{\bf EXO 0748-676} & 117.149 & -67.754 & 0.8 & 21.7$\pm$0.6 & 18.1$\pm$1.0 & LMXB,B,D,T &  43.7 & 116.0& B5 \\ 
IGR J07565-4139        & 119.110 & -41.631 & 3.8 &  1.2$\pm$0.2 &  0.9$\pm$0.3 & AGN,Sy2 &   6.8 & 968.0& B4 \\ 
IGR J07597-3842        & 119.923 & -38.719 & 2.3 &  2.2$\pm$0.2 &  2.3$\pm$0.3 & AGN,Sy1.2 &  11.8 & 837.0& B5 \\ 
ESO 209-12        & 120.507 & -49.753 & 3.6 &  0.9$\pm$0.2 &  1.7$\pm$0.3 & AGN,Sy1.5 &   7.2 & 936.0& B5 \\ 
{\bf QSO B0804+761} & 122.781 & 76.010 & 5.2 &  0.9$\pm$0.2 & $<$ 0.9 & AGN,Sy1 &   4.7 & 528.0& B5 \\ 
Vela Pulsar        & 128.831 & -45.182 & 0.6 &  7.1$\pm$0.1 &  8.1$\pm$0.2 & PWN,PSR &  65.5 & 1556.0& B5 \\ 
4U 0836-429        & 129.346 & -42.894 & 0.2 & 31.2$\pm$0.1 & 26.6$\pm$0.2 & LMXB,B,T & 378.5 & 1576.0& S137B3 \\ 
FRL 1146        & 129.620 & -36.013 & 3.5 &  1.2$\pm$0.2 &  0.9$\pm$0.3 & AGN,Sy1.5 &   7.4 & 1041.0& B5 \\ 
{\bf IGR J08408-4503} & 130.203 & -45.061 & 5.3 & $<$ 0.2 &  0.5$\pm$0.2 & HMXB,SFXT &   4.6 & 1643.0& B3 \\ 
{\bf QSO B0836+710} & 130.296 & 70.897 & 2.0 &  2.6$\pm$0.2 &  3.8$\pm$0.4 & AGN,Blazar &  13.4 & 506.0& B5 \\ 
Vela X-1        & 135.523 & -40.554 & 0.2 & 221.0$\pm$0.1 & 51.1$\pm$0.2 & HMXB,XP & 1554.0 & 1527.0& B1 \\ 
{\bf IGR J09025-6814} & 135.612 & -68.235 & 5.1 &  1.2$\pm$0.3 &  1.7$\pm$0.5 & ? &   4.8 & 381.0& S192B3 \\ 
IGR J09026-4812        & 135.668 & -48.216 & 2.3 &  1.3$\pm$0.1 &  1.4$\pm$0.2 & ? &  11.5 & 1527.0& B5 \\ 
{\bf IGR J09103-3741} & 137.577 & -37.675 & 5.2 & $<$ 0.3 & $<$ 0.5 & ? &   4.7 & 965.0& S315B1 \\ 
{\bf SWIFT J0917.2-6221} & 139.040 & -62.314 & 3.0 &  1.5$\pm$0.2 &  1.1$\pm$0.3 & AGN,Sy1 &   8.7 & 998.0& B5 \\ 
EXMS B0918-549E        & 140.051 & -55.125 & 4.1 &  3.0$\pm$0.2 &  2.2$\pm$0.3 & ?,T &   6.2 & 1317.0& R139B1 \\ 
4U 0919-54        & 140.093 & -55.191 & 1.1 &  4.0$\pm$0.2 &  3.0$\pm$0.3 & LMXB &  26.4 & 1306.0& B5 \\ 
{\bf IGR J09253+6929} & 141.320 & 69.488 & 4.7 &  1.4$\pm$0.3 & $<$ 1.2 & ? &   5.3 & 290.0& B5 \\ 
{\bf IGR J09469-4603} & 146.722 & -46.046 & 4.8 &  0.7$\pm$0.2 &  0.9$\pm$0.3 & ? &   5.1 & 1199.0& B3 \\ 
MCG-05-23-016        & 146.916 & -30.936 & 1.9 &  9.2$\pm$0.6 &  8.0$\pm$1.0 & AGN,Sy2 &  15.0 &  90.0& B5 \\ 
{\bf IGR J09485-4726} & 147.120 & -47.427 & 4.9 &  0.5$\pm$0.2 &  1.3$\pm$0.3 & ? &   5.1 & 1158.0& B3 \\ 
{\bf IGR J09523-6231} & 148.069 & -62.516 & 3.9 &  0.9$\pm$0.1 &  0.6$\pm$0.3 & ? &   6.4 & 1227.0& B5 \\ 
{\bf SWIFT J1009.3-4250} & 152.393 & -42.784 & 3.3 &  1.8$\pm$0.3 &  1.6$\pm$0.4 & AGN,Sy2 &   7.8 & 600.0& B5 \\ 
GRO J1008-57        & 152.434 & -58.296 & 0.5 &  4.6$\pm$0.1 &  2.1$\pm$0.2 & HMXB,XP,Be,T &  74.5 & 1571.0& R203B1 \\ 
{\bf IGR J10101-5654} & 152.523 & -56.916 & 2.8 &  1.1$\pm$0.1 &  0.6$\pm$0.2 & HMXB &   9.3 & 1520.0& B5 \\ 
{\bf IGR J10109-5746} & 152.666 & -57.788 & 2.7 &  1.1$\pm$0.1 & $<$ 0.4 & Symb &   9.7 & 1585.0& B4 \\ 
{\bf IGR J10147-6354} & 153.677 & -63.892 & 5.0 & $<$ 0.3 &  1.3$\pm$0.2 & ? &   4.9 & 1340.0& B3 \\ 
{\bf NGC 3281} & 157.951 & -34.869 & 3.9 &  2.3$\pm$0.5 &  3.6$\pm$0.9 & AGN,Sy2 &   6.5 & 110.0& B5 \\ 
{\bf 4U 1036-56} & 159.373 & -56.784 & 3.5 &  0.8$\pm$0.1 & $<$ 0.4 & HMXB,Be &   7.3 & 1555.0& S081B3 \\ 
{\bf SWIFT J1038.8-4942} & 159.652 & -49.829 & 4.0 &  0.8$\pm$0.2 &  0.9$\pm$0.3 & AGN,Sy1.5 &   6.3 & 1016.0& B5 \\ 
IGR J10404-4625        & 160.107 & -46.410 & 3.1 &  1.9$\pm$0.2 &  1.9$\pm$0.4 & AGN,Sy2 &   8.3 & 626.0& B3 \\ 
{\bf IGR J10448-5945\tablenotemark{k}} & 161.197 & -59.755 & 5.2 &  0.3$\pm$0.1 &  0.9$\pm$0.2 & ? &   4.8 & 1578.0& B3 \\ 
{\bf IGR J11098-6457} & 167.442 & -64.946 & 5.5 &  0.6$\pm$0.2 &  0.9$\pm$0.2 & ? &   4.5 & 1252.0& B5 \\ 
{\bf IGR J11187-5438} & 169.677 & -54.633 & 4.0 &  1.0$\pm$0.1 &  0.8$\pm$0.2 & ? &   6.3 & 1269.0& B1 \\ 
Cen X-3        & 170.307 & -60.627 & 0.2 & 64.3$\pm$0.1 &  7.2$\pm$0.2 & HMXB,XP & 542.9 & 1445.0& B4 \\ 
{\bf IGR J11215-5952} & 170.450 & -59.869 & 1.5 &  0.5$\pm$0.1 & $<$ 0.5 & HMXB,SFXT &  19.1 & 1422.0& S308B1 \\ 
{\bf IGR J11305-6256} & 172.775 & -62.939 & 1.2 &  3.9$\pm$0.1 &  1.7$\pm$0.2 & HMXB,Be &  26.1 & 1337.0& B5 \\ 
{\bf IGR J11366-6002} & 174.158 & -60.034 & 4.9 &  0.6$\pm$0.1 &  0.9$\pm$0.2 & AGN? &   5.0 & 1302.0& B3 \\ 
{\bf NGC 3783} & 174.733 & -37.745 & 3.2 & 10.5$\pm$1.2 &  5.8$\pm$2.2 & AGN,Sy1 &   8.1 &  23.0& B1 \\ 
{\bf EXMS B1136-650} & 174.870 & -65.405 & 3.2 & $<$ 0.3 & $<$ 0.5 & RSCVn &   8.0 & 1203.0& R088B1 \\ 
{\bf IGR J11435-6109} & 175.969 & -61.145 & 2.6 &  1.2$\pm$0.2 &  0.9$\pm$0.2 & HMXB,XP?,Be &  10.4 & 1292.0& R258B2 \\ 
1E 1145.1-6141        & 176.864 & -61.963 & 0.3 & 26.7$\pm$0.2 & 15.7$\pm$0.3 & HMXB,XP & 172.2 & 1305.0& B5 \\ 
2E 1145.5-6155        & 176.981 & -62.187 & 1.0 &  4.0$\pm$0.2 &  2.4$\pm$0.3 & HMXB,XP &  31.6 & 1282.0& R076B2 \\ 
{\bf IGR J12026-5349} & 180.642 & -53.849 & 2.2 &  2.5$\pm$0.2 &  2.2$\pm$0.3 & AGN,Sy2 &  12.3 & 728.0& B3 \\ 
{\bf NGC 4138} & 182.393 & 43.676 & 5.3 &  2.5$\pm$0.6 &  2.9$\pm$1.1 & AGN,Sy1.9 &   4.6 &  62.0& B5 \\ 
NGC 4151        & 182.636 & 39.409 & 0.4 & 33.7$\pm$0.6 & 40.2$\pm$1.0 & AGN,Sy1.5 & 144.0 &  71.0& STB1 \\ 
{\bf EXMS B1210-645} & 183.272 & -64.897 & 4.9 &  0.7$\pm$0.2 &  0.6$\pm$0.2 & ?,T &   5.0 & 1168.0& B5 \\ 
{\bf NGC 4180\tablenotemark{l}} & 183.291 &  7.009 & 5.2 &  0.6$\pm$0.2 &  1.6$\pm$0.4 & AGN &   4.7 & 622.0& B3 \\ 
4C 04.42        & 185.612 &  4.256 & 3.2 &  1.0$\pm$0.2 &  2.0$\pm$0.3 & AGN,QSO &   8.0 & 810.0& B3 \\ 
{\bf Mrk 50} & 185.862 &  2.693 & 3.4 &  1.1$\pm$0.2 &  0.6$\pm$0.3 & AGN,Sy1 &   7.6 & 848.0& B4 \\ 
{\bf NGC 4395} & 186.364 & 33.556 & 5.4 &  1.5$\pm$0.4 &  1.8$\pm$0.7 & AGN,Sy1.8 &   4.6 & 144.0& B5 \\ 
NGC 4388        & 186.445 & 12.658 & 0.7 & 12.4$\pm$0.3 & 15.2$\pm$0.5 & AGN,Sy2 &  53.1 & 411.0& B5 \\ 
GX 301-2        & 186.649 & -62.772 & 0.2 & 134.0$\pm$0.2 & 17.9$\pm$0.3 & HMXB,XP,T & 786.1 & 1038.0& B4 \\ 
{\bf XSS J12270-4859} & 187.007 & -48.893 & 4.3 &  1.8$\pm$0.3 &  1.7$\pm$0.5 & CV,IP &   5.9 & 336.0& B5 \\ 
3C 273        & 187.280 &  2.049 & 0.6 & 10.1$\pm$0.2 & 11.5$\pm$0.3 & AGN,QSO &  69.0 & 892.0& B5 \\ 
IGR J12349-6434        & 188.722 & -64.570 & 1.2 &  4.4$\pm$0.2 &  3.3$\pm$0.3 & Symb &  24.6 & 965.0& B5 \\ 
NGC 4507        & 188.908 & -39.904 & 1.1 &  8.9$\pm$0.4 & 11.3$\pm$0.7 & AGN,Sy2 &  27.6 & 215.0& B5 \\ 
{\bf ESO 506-G27} & 189.748 & -27.237 & 5.1 &  4.8$\pm$1.1 & $<$ 0.3 & AGN,Sy2 &   4.9 &  28.0& B5 \\ 
LEDA 170194        & 189.778 & -16.190 & 3.3 &  2.4$\pm$0.4 &  4.6$\pm$0.7 & AGN,Sy2 &   7.8 & 200.0& BÂ3 \\ 
NGC 4593        & 189.905 & -5.353 & 1.1 &  4.2$\pm$0.2 &  4.3$\pm$0.3 & AGN,Sy1 &  27.0 & 817.0& B5 \\ 
{\bf IGR J12415-5750} & 190.377 & -57.825 & 3.7 &  1.4$\pm$0.2 &  1.3$\pm$0.3 & AGN,Sy2 &   6.8 & 727.0& B5 \\ 
{\bf 1H 1249-637} & 190.725 & -63.049 & 4.0 &  1.1$\pm$0.2 & $<$ 0.6 & HMXB,Be &   6.3 & 963.0& B4 \\ 
4U 1246-58        & 192.410 & -59.090 & 1.6 &  3.1$\pm$0.2 &  2.4$\pm$0.3 & LMXB,B &  17.2 & 848.0& B5 \\ 
{\bf ESO 323-32} & 193.394 & -41.626 & 4.7 &  0.9$\pm$0.3 &  1.6$\pm$0.5 & AGN,Sy2 &   5.3 & 322.0& B5 \\ 
3C 279        & 194.037 & -5.742 & 3.5 &  0.9$\pm$0.2 &  1.8$\pm$0.3 & AGN,Blazar &   7.3 & 685.0& B3 \\ 
1H 1254-690        & 194.438 & -69.288 & 1.6 &  2.5$\pm$0.2 & $<$ 0.7 & LMXB,B,D &  17.2 & 711.0& B4 \\ 
Coma Cluster        & 194.884 & 27.939 & 2.9 &  1.8$\pm$0.3 & $<$ 1.0 & Cluster &   9.1 & 286.0& B4 \\ 
IGR J13020-6359        & 195.537 & -63.947 & 2.3 &  2.1$\pm$0.2 &  1.4$\pm$0.3 & HMXB,XP,Be &  11.5 & 926.0& B5 \\ 
NGC 4945        & 196.361 & -49.470 & 0.6 & 13.7$\pm$0.2 & 20.7$\pm$0.4 & AGN,Sy2 &  69.4 & 511.0& B3 \\ 
{\bf ESO 323-77} & 196.621 & -40.449 & 4.2 &  1.2$\pm$0.3 &  2.1$\pm$0.5 & AGN,Sy1.2 &   6.0 & 346.0& B3 \\ 
{\bf IGR J13091+1137} & 197.290 & 11.635 & 3.7 &  2.4$\pm$0.4 &  2.9$\pm$0.6 & AGN,Sy2,XBONG &   6.8 & 268.0& B3 \\ 
{\bf IGR J13109-5552} & 197.682 & -55.863 & 3.4 &  1.2$\pm$0.2 &  1.6$\pm$0.3 & AGN? &   7.4 & 762.0& B3 \\ 
{\bf NGC 5033} & 198.348 & 36.571 & 4.9 &  1.3$\pm$0.3 & $<$ 1.2 & AGN,Sy1.9 &   5.1 & 189.0& B5 \\ 
{\bf IGR J13186-6257} & 199.652 & -62.946 & 3.8 &  0.8$\pm$0.2 & $<$ 0.6 & ? &   6.6 & 958.0& B5 \\ 
Cen A        & 201.365 & -43.020 & 0.3 & 38.1$\pm$0.2 & 48.4$\pm$0.4 & AGN,Sy2 & 183.3 & 421.0& B5 \\ 
4U 1323-62        & 201.634 & -62.136 & 0.9 &  5.9$\pm$0.2 &  3.5$\pm$0.3 & LMXB,B,D &  35.7 & 992.0& B5 \\ 
{\bf 1RXS J133447.5+371100} & 203.793 & 37.199 & 4.2 &  2.1$\pm$0.4 & $<$ 1.5 & XB &   6.0 & 139.0& B4 \\ 
{\bf MCG-06-30-015} & 203.995 & -34.302 & 2.1 &  3.7$\pm$0.3 &  2.0$\pm$0.5 & AGN,Sy1.2 &  13.1 & 325.0& B1 \\ 
{\bf NGC 5252} & 204.559 &  4.504 & 4.6 &  3.5$\pm$0.6 & $<$ 2.1 & AGN,Sy1.9 &   5.5 & 120.0& B1 \\ 
4U 1344-60        & 206.883 & -60.610 & 1.0 &  4.3$\pm$0.2 &  4.4$\pm$0.3 & AGN,Sy1.5 &  29.5 & 1042.0& B5 \\ 
IC 4329A        & 207.339 & -30.309 & 0.8 & 12.0$\pm$0.3 & 13.2$\pm$0.6 & AGN,Sy1.2 &  41.7 & 239.0& B5 \\ 
{\bf IGR J14003-6326} & 210.154 & -63.447 & 3.9 &  1.1$\pm$0.2 &  1.0$\pm$0.3 & ? &   6.5 & 1052.0& B1 \\ 
{\bf V834 Cen} & 212.249 & -45.273 & 4.4 &  1.1$\pm$0.2 & $<$ 0.7 & CV,P &   5.7 & 726.0& B4 \\ 
Circinus Galaxy        & 213.282 & -65.345 & 0.5 & 14.1$\pm$0.2 & 11.7$\pm$0.3 & AGN,Sy2 &  88.4 & 991.0& B5 \\ 
{\bf NGC 5506} & 213.318 & -3.197 & 2.1 &  6.3$\pm$0.5 &  4.3$\pm$0.9 & AGN,Sy1.9 &  12.9 &  71.0& STB1 \\ 
{\bf IGR J14175-4641} & 214.275 & -46.676 & 5.5 &  0.5$\pm$0.2 &  1.1$\pm$0.4 & AGN,Sy2 &   4.5 & 790.0& B3 \\ 
{\bf ESO 511-G030} & 214.860 & -26.644 & 4.0 &  2.2$\pm$0.4 &  1.9$\pm$0.8 & AGN,Sy1 &   6.3 & 180.0& B5 \\ 
{\bf IGR J14298-6715} & 217.338 & -67.260 & 4.3 &  0.7$\pm$0.2 &  0.8$\pm$0.3 & ? &   5.9 & 856.0& B5 \\ 
{\bf IGR J14319-3315} & 217.988 & -33.245 & 4.8 &  1.2$\pm$0.3 & $<$ 0.9 & ? &   5.1 & 449.0& B1 \\ 
{\bf IGR J14331-6112} & 218.357 & -61.204 & 3.7 &  0.8$\pm$0.2 &  0.7$\pm$0.3 & ? &   6.8 & 1216.0& B5 \\ 
{\bf IGR J14471-6414\tablenotemark{m}} & 221.588 & -64.294 & 4.6 &  0.6$\pm$0.2 &  0.8$\pm$0.3 & ? &   5.4 & 1066.0& B5 \\ 
{\bf IGR J14471-6319} & 221.872 & -63.309 & 4.6 &  0.7$\pm$0.2 &  1.0$\pm$0.3 & AGN,Sy2 &   5.5 & 1115.0& B5 \\ 
IGR J14492-5535        & 222.358 & -55.561 & 3.3 &  1.2$\pm$0.1 &  0.7$\pm$0.3 & AGN &   7.8 & 1427.0& B1 \\ 
{\bf IGR J14515-5542} & 222.904 & -55.669 & 3.4 &  0.9$\pm$0.1 &  1.0$\pm$0.3 & AGN,Sy2 &   7.6 & 1453.0& B5 \\ 
{\bf IGR J14532-6356} & 223.312 & -63.927 & 5.3 & $<$ 0.3 & $<$ 0.6 & ? &   4.6 & 1099.0& R036B1 \\ 
{\bf IGR J14536-5522} & 223.435 & -55.374 & 2.7 &  1.5$\pm$0.1 & $<$ 0.5 & CV &   9.9 & 1475.0& B1 \\ 
{\bf IGR J14552-5133} & 223.792 & -51.588 & 4.4 &  0.8$\pm$0.2 &  0.6$\pm$0.3 & AGN,NL Sy1 &   5.6 & 1375.0& B5 \\ 
{\bf IC 4518A} & 224.391 & -43.156 & 2.3 &  1.6$\pm$0.2 &  1.1$\pm$0.3 & AGN,Sy2 &  11.5 & 898.0& B5 \\ 
{\bf IGR J15094-6649} & 227.351 & -66.844 & 2.9 &  1.7$\pm$0.2 & $<$ 0.7 & CV,IP &   8.9 & 833.0& B1 \\ 
PSR B1509-58        & 228.486 & -59.147 & 0.5 &  9.0$\pm$0.1 & 10.9$\pm$0.2 & PSR &  70.6 & 1439.0& B5 \\ 
{\bf ESO 328-IG036} & 228.755 & -40.380 & 4.8 &  0.8$\pm$0.2 &  0.6$\pm$0.3 & AGN,Sy1 &   5.2 & 931.0& B5 \\ 
{\bf IGR J15161-3827} & 229.037 & -38.448 & 5.3 &  0.5$\pm$0.2 &  1.2$\pm$0.3 & AGN? &   4.7 & 816.0& B2 \\ 
Cir X-1        & 230.172 & -57.169 & 0.4 &  9.3$\pm$0.1 &  0.6$\pm$0.2 & LMXB,B,A,T &  99.1 & 1557.0& B4 \\ 
IGR J15359-5750        & 234.031 & -57.803 & 2.5 &  1.1$\pm$0.1 &  1.5$\pm$0.2 & ? &  10.5 & 1663.0& B5 \\ 
4U 1538-522        & 235.595 & -52.388 & 0.3 & 22.3$\pm$0.1 &  3.4$\pm$0.2 & HMXB,XP & 162.8 & 1999.0& B1 \\ 
{\bf XTE J1543-568} & 236.010 & -56.712 & 4.2 &  0.5$\pm$0.1 & $<$ 0.5 & HMXB,XP,Be,T &   6.0 & 1771.0& S047B1 \\ 
4U 1543-624        & 236.976 & -62.570 & 1.3 &  3.4$\pm$0.2 &  0.6$\pm$0.3 & LMXB,NS? &  23.0 & 1126.0& B4 \\ 
IGR J15479-4529        & 237.050 & -45.478 & 0.9 &  5.1$\pm$0.1 &  3.1$\pm$0.2 & CV,IP &  35.8 & 1622.0& B5 \\ 
{\bf NGC 5995} & 237.113 & -13.762 & 3.8 &  2.2$\pm$0.6 &  2.1$\pm$0.5 & AGN,Sy2 &   6.7 & 138.0& B2 \\ 
XTE J1550-564        & 237.745 & -56.479 & 0.2 & 34.0$\pm$0.1 & 55.3$\pm$0.2 & LMXB,BH,T & 737.9 & 1821.0& S047B3 \\ 
{\bf IGR J15529-5029} & 238.235 & -50.490 & 3.9 &  0.9$\pm$0.1 & $<$ 0.4 & ? &   6.5 & 2020.0& B5 \\ 
{\bf IGR J15539-6142} & 238.336 & -61.671 & 3.9 &  0.7$\pm$0.2 &  1.7$\pm$0.3 & AGN? &   6.4 & 1220.0& B3 \\ 
{\bf 1H 1556-605} & 240.312 & -60.754 & 2.9 &  0.8$\pm$0.2 &  0.6$\pm$0.3 & LMXB &   9.1 & 1305.0& B4 \\ 
{\bf IGR J16024-6107} & 240.609 & -61.124 & 4.7 &  0.9$\pm$0.2 & $<$ 0.5 & AGN? &   5.2 & 1254.0& B1 \\ 
{\bf IGR J16056-6110} & 241.394 & -61.171 & 5.0 &  0.5$\pm$0.2 &  1.3$\pm$0.3 & AGN &   4.9 & 1229.0& B3 \\ 
{\bf IGR J16119-6036} & 242.988 & -60.658 & 2.6 &  1.5$\pm$0.2 &  1.5$\pm$0.3 & AGN,Sy1 &  10.2 & 1263.0& B5 \\ 
4U 1608-522        & 243.177 & -52.424 & 0.4 & 13.9$\pm$0.1 &  7.8$\pm$0.2 & LMXB,B,A,T & 119.8 & 1984.0& B4 \\ 
IGR J16167-4957        & 244.140 & -49.974 & 1.7 &  2.1$\pm$0.1 &  0.7$\pm$0.2 & CV,IP &  16.3 & 2002.0& B4 \\ 
{\bf PSR J1617-5055} & 244.303 & -50.942 & 3.5 &  0.9$\pm$0.1 &  0.9$\pm$0.2 & PSR &   7.4 & 2007.0& B5 \\ 
{\bf IGR J16185-5928} & 244.607 & -59.446 & 3.8 &  1.2$\pm$0.2 &  1.0$\pm$0.3 & AGN,NL Sy1 &   6.8 & 1314.0& B5 \\ 
AX J1619.4-4945        & 244.895 & -49.744 & 1.8 &  2.0$\pm$0.1 &  1.2$\pm$0.2 & HMXB?,SFXT? &  15.6 & 1964.0& B5 \\ 
IGR J16194-2810        & 244.899 & -28.138 & 3.5 &  2.5$\pm$0.3 &  1.2$\pm$0.4 & ? &   7.4 & 461.0& B1 \\ 
Sco X-1        & 244.980 & -15.643 & 0.2 & 685.7$\pm$0.3 & 24.7$\pm$0.3 & LMXB,Z & 2422.7 & 423.0& B4 \\ 
IGR J16207-5129        & 245.189 & -51.505 & 1.2 &  3.3$\pm$0.1 &  2.3$\pm$0.2 & HMXB,SG &  26.0 & 1943.0& B5 \\ 
{\bf IGR J16248-4603} & 246.207 & -46.043 & 4.7 & $<$ 0.3 & $<$ 0.4 & ? &   5.3 & 1864.0& S163B2 \\ 
{\bf SWIFT J1626.6-5156} & 246.648 & -51.944 & 2.0 & $<$ 0.3 &  0.4$\pm$0.2 & LMXB?,XP,T &  14.0 & 1884.0& R399B2 \\ 
4U 1624-490        & 247.013 & -49.208 & 0.7 &  4.4$\pm$0.1 &  0.4$\pm$0.2 & LMXB,D &  48.9 & 1933.0& B4 \\ 
{\bf IGR J16283-4838} & 247.041 & -48.644 & 3.3 &  0.7$\pm$0.1 &  0.5$\pm$0.2 & HMXB?,NS? &   7.8 & 1896.0& R303B2 \\ 
{\bf IGR J16287-5021} & 247.174 & -50.344 & 4.4 &  0.8$\pm$0.1 & $<$ 0.4 & ? &   5.7 & 1916.0& B1 \\ 
IGR J16318-4848        & 247.952 & -48.820 & 0.3 & 24.8$\pm$0.1 & 12.8$\pm$0.2 & HMXB & 179.8 & 1945.0& B5 \\ 
AX J1631.9-4752        & 248.006 & -47.875 & 0.4 & 17.4$\pm$0.1 &  6.4$\pm$0.2 & HMXB,XP,T & 121.7 & 1830.0& B1 \\ 
4U 1626-67        & 248.082 & -67.468 & 0.6 & 18.4$\pm$0.3 &  1.8$\pm$0.5 & LMXB,XP &  65.3 & 445.0& B4 \\ 
{\bf IGR J16328-4726} & 248.190 & -47.437 & 4.5 &  4.0$\pm$0.1 &  3.2$\pm$0.2 & ? &   5.6 & 1718.0& S351B3 \\ 
4U 1630-47        & 248.507 & -47.396 & 0.2 & 38.9$\pm$0.1 & 32.3$\pm$0.2 & LMXB,BHC,D,T & 407.5 & 1842.0& S100B1 \\ 
{\bf IGR J16351-5806} & 248.796 & -58.090 & 3.1 &  1.0$\pm$0.2 &  1.3$\pm$0.3 & AGN,Sy2 &   8.3 & 1397.0& B5 \\ 
IGR J16358-4726        & 248.976 & -47.425 & 0.8 &  2.1$\pm$0.1 &  1.1$\pm$0.2 & ?,XP,T &  40.3 & 1697.0& R052B1 \\ 
IGR J16377-6423        & 249.557 & -64.356 & 3.7 &  1.3$\pm$0.2 & $<$ 0.8 & Cluster? &   6.9 & 649.0& B4 \\ 
{\bf IGR J16385-2057} & 249.626 & -20.944 & 4.1 &  1.4$\pm$0.2 &  0.6$\pm$0.3 & AGN &   6.2 & 726.0& B3 \\ 
AX J1639.0-4642        & 249.775 & -46.706 & 0.7 &  6.3$\pm$0.1 &  0.7$\pm$0.2 & HMXB,XP,T &  48.7 & 1890.0& B4 \\ 
4U 1636-536        & 250.228 & -53.753 & 0.3 & 38.1$\pm$0.1 & 23.4$\pm$0.2 & LMXB,B,A & 272.1 & 1713.0& B5 \\ 
IGR J16418-4532        & 250.468 & -45.548 & 1.0 &  4.7$\pm$0.1 &  1.1$\pm$0.2 & HMXB,XP,SFXT? &  30.7 & 1847.0& B1 \\ 
{\bf IGR J16426+6536} & 250.656 & 65.594 & 4.5 &  3.1$\pm$0.6 & $<$ 2.1 & ? &   5.5 &  97.0& R206B1 \\ 
GX 340+0        & 251.448 & -45.614 & 0.2 & 31.9$\pm$0.1 &  1.5$\pm$0.2 & LMXB,Z & 332.6 & 1920.0& B4 \\ 
{\bf IGR J16460+0849} & 251.489 &  8.818 & 5.2 &  6.8$\pm$2.4 & 12.6$\pm$3.8 & ? &   4.7 &  11.0& B3 \\ 
{\bf IGR J16465-4507} & 251.696 & -45.125 & 2.1 &  1.8$\pm$0.1 &  1.1$\pm$0.2 & HMXB,SFXT?,XP &  12.8 & 1916.0& R222B2 \\ 
IGR J16479-4514        & 252.015 & -45.216 & 1.4 &  4.5$\pm$0.1 &  2.4$\pm$0.2 & HMXB,SFXT? &  20.2 & 1950.0& B2 \\ 
IGR J16482-3036        & 252.062 & -30.590 & 2.2 &  1.7$\pm$0.2 &  1.9$\pm$0.2 & AGN,Sy1 &  12.4 & 1723.0& B3 \\ 
IGR J16493-4348        & 252.375 & -43.828 & 1.7 &  2.3$\pm$0.1 &  1.6$\pm$0.2 & LMXB? &  16.8 & 2053.0& B5 \\ 
IGR J16500-3307        & 252.491 & -33.064 & 2.4 &  1.8$\pm$0.2 &  0.5$\pm$0.2 & ? &  11.3 & 1943.0& B1 \\ 
ESO 138-1\tablenotemark{n}        & 253.049 & -59.222 & 3.8 &  1.3$\pm$0.2 &  1.1$\pm$0.3 & AGN,Sy2 &   6.8 & 1066.0& B5 \\ 
NGC 6221\tablenotemark{o}        & 253.049 & -59.222 & 3.8 &  1.3$\pm$0.2 &  1.1$\pm$0.3 & AGN,Sy1/Sy2 &   6.8 & 1066.0& B5 \\ 
{\bf GRO J1655-40} & 253.504 & -39.846 & 0.6 &  2.3$\pm$0.1 &  2.7$\pm$0.2 & LMXB,BH,T &  58.9 & 2406.0& S296B3 \\ 
IGR J16558-5203        & 254.010 & -52.062 & 2.0 &  1.8$\pm$0.1 &  2.1$\pm$0.2 & AGN,Sy1.2 &  13.9 & 1509.0& B3 \\ 
{\bf Swift J1656.3-3302} & 254.110 & -33.047 & 2.5 &  1.2$\pm$0.1 &  1.5$\pm$0.2 & AGN? &  10.6 & 2314.0& B3 \\ 
{\bf Her X-1} & 254.461 & 35.339 & 2.7 & 87.0$\pm$9.0 & $<$ 20.0 & LMXB,XP &   9.6 &   2.0& STB1 \\ 
AX J1700.2-4220        & 255.059 & -42.308 & 2.0 &  1.9$\pm$0.1 &  1.5$\pm$0.2 & HMXB &  14.1 & 1991.0& B5 \\ 
OAO 1657-415        & 255.203 & -41.659 & 0.2 & 76.7$\pm$0.1 & 40.4$\pm$0.2 & HMXB,XP & 529.1 & 2223.0& B5 \\ 
{\bf IGR J17008-6425} & 255.204 & -64.425 & 4.5 &  1.2$\pm$0.3 &  1.6$\pm$0.5 & ? &   5.6 & 499.0& B5 \\ 
{\bf XTE J1701-462} & 255.232 & -46.182 & 1.1 &  0.6$\pm$0.1 & $<$ 0.4 & LMXB,Z,T &  26.3 & 1761.0& R411B2 \\ 
GX 339-4        & 255.706 & -48.792 & 0.3 & 40.7$\pm$0.1 & 46.7$\pm$0.2 & LMXB,BH,T & 306.3 & 1590.0& B3 \\ 
4U 1700-377        & 255.987 & -37.847 & 0.2 & 208.1$\pm$0.1 & 123.8$\pm$0.2 & HMXB & 1670.4 & 3106.0& B5 \\ 
GX 349+2        & 256.441 & -36.426 & 0.2 & 48.7$\pm$0.1 &  1.1$\pm$0.2 & LMXB,Z & 574.5 & 3120.0& B4 \\ 
4U 1702-429        & 256.559 & -43.041 & 0.4 & 15.2$\pm$0.1 &  9.1$\pm$0.2 & LMXB,B,A & 108.5 & 2089.0& B5 \\ 
{\bf 1H 1705-250} & 257.065 & -25.094 & 5.0 &  0.5$\pm$0.1 & $<$ 0.3 & LMXB,BHC,T &   5.0 & 2555.0& S411B3 \\ 
IGR J17088-4008        & 257.220 & -40.164 & 2.3 &  1.2$\pm$0.1 &  2.4$\pm$0.2 & AXP &  11.8 & 2665.0& B3 \\ 
4U 1705-32        & 257.221 & -32.315 & 1.2 &  2.8$\pm$0.1 &  2.7$\pm$0.2 & LMXB,B &  24.4 & 3340.0& B3 \\ 
4U 1705-440        & 257.223 & -44.105 & 0.3 & 28.3$\pm$0.1 & 13.1$\pm$0.2 & LMXB,B,A & 207.1 & 1940.0& B4 \\ 
IGR J17091-3624        & 257.280 & -36.407 & 0.5 &  6.7$\pm$0.1 &  8.9$\pm$0.2 & XB,BHC? &  74.6 & 3134.0& S163B3 \\ 
XTE J1709-267        & 257.393 & -26.651 & 1.8 &  0.8$\pm$0.1 &  0.7$\pm$0.2 & LMXB,B,T &  15.9 & 2758.0& R171B1 \\ 
{\bf IGR J17098-3628} & 257.462 & -36.463 & 1.9 &  2.9$\pm$0.1 &  2.9$\pm$0.2 & ?,BHC?,T &  14.4 & 3028.0& S301B3 \\ 
XTE J1710-281        & 257.539 & -28.123 & 1.1 &  3.0$\pm$0.1 &  3.0$\pm$0.2 & LMXB,B,T &  29.1 & 3040.0& B3 \\ 
4U 1708-40        & 258.104 & -40.859 & 2.3 &  1.1$\pm$0.1 &  0.6$\pm$0.2 & LMXB,B &  12.0 & 2534.0& B4 \\ 
Oph Cluster        & 258.112 & -23.348 & 0.8 &  5.2$\pm$0.1 &  1.1$\pm$0.2 & Cluster &  41.9 & 2869.0& B4 \\ 
SAX J1712.6-3739        & 258.137 & -37.645 & 0.8 &  4.7$\pm$0.1 &  4.0$\pm$0.2 & LMXB,B,T &  41.0 & 3263.0& B5 \\ 
V2400 Oph        & 258.172 & -24.280 & 1.2 &  3.5$\pm$0.1 &  1.0$\pm$0.2 & CV,IP &  24.3 & 2765.0& B1 \\ 
XTE J1716-389        & 258.915 & -38.875 & 3.5 &  1.0$\pm$0.1 &  0.6$\pm$0.2 & HMXB,SG &   7.4 & 3003.0& B4 \\ 
NGC 6300        & 259.229 & -62.822 & 1.8 &  4.4$\pm$0.3 &  3.6$\pm$0.5 & AGN,Sy2 &  15.2 & 494.0& B5 \\ 
IGR J17195-4100        & 259.906 & -41.014 & 1.7 &  2.4$\pm$0.1 &  1.5$\pm$0.2 & CV,IP &  16.9 & 2531.0& B5 \\ 
XTE J1720-318        & 259.995 & -31.760 & 0.7 &  1.7$\pm$0.1 &  2.7$\pm$0.1 & LMXB,BHC,T &  53.8 & 4127.0& S047B3 \\ 
IGR J17200-3116        & 260.025 & -31.284 & 1.1 &  2.8$\pm$0.1 &  1.2$\pm$0.1 & HMXB,T &  26.5 & 4192.0& B5 \\ 
IGR J17204-3554        & 260.104 & -35.892 & 2.0 &  1.1$\pm$0.1 &  2.0$\pm$0.2 & AGN? &  14.0 & 3620.0& B3 \\ 
EXO 1722-363        & 261.299 & -36.282 & 0.5 & 10.3$\pm$0.1 &  3.0$\pm$0.2 & HMXB,XP &  91.8 & 3744.0& B1 \\ 
IGR J17254-3257        & 261.361 & -32.971 & 1.3 &  2.2$\pm$0.1 &  2.1$\pm$0.1 & LMXB,B &  23.2 & 4513.0& B5 \\ 
{\bf IGR J17269-4737} & 261.755 & -47.622 & 3.7 & $<$ 0.3 &  0.5$\pm$0.3 & ?,T &   6.8 & 1285.0& S358B1 \\ 
GRS 1724-308        & 261.891 & -30.805 & 0.3 & 19.4$\pm$0.1 & 15.7$\pm$0.1 & LMXB,G,B,A & 223.8 & 4626.0& B5 \\ 
IGR J17285-2922        & 262.143 & -29.375 & 3.5 & $<$ 0.2 & $<$ 0.3 & XB?,T &   7.2 & 4381.0& R120B1 \\ 
IGR J17303-0601        & 262.596 & -5.988 & 1.9 &  3.8$\pm$0.3 &  2.7$\pm$0.4 & CV,IP &  14.7 & 498.0& B5 \\ 
{\bf IGR J17314-2854} & 262.853 & -28.895 & 3.2 &  0.3$\pm$0.1 &  0.5$\pm$0.1 & ? &   8.0 & 4775.0& S100B3 \\ 
GX 9+9        & 262.933 & -16.963 & 0.4 & 14.1$\pm$0.1 &  0.5$\pm$0.2 & LMXB,A & 141.5 & 2098.0& B4 \\ 
{\bf V2487 Oph} & 262.963 & -19.233 & 3.1 &  0.8$\pm$0.1 &  0.8$\pm$0.2 & CV,IP? &   8.5 & 2635.0& B5 \\ 
GX 354-0        & 262.991 & -33.834 & 0.2 & 41.1$\pm$0.1 & 14.1$\pm$0.1 & LMXB,B,A & 471.6 & 4550.0& B4 \\ 
GX 1+4        & 263.008 & -24.747 & 0.2 & 50.6$\pm$0.1 & 39.1$\pm$0.1 & LMXB,XP & 544.3 & 4161.0& B5 \\ 
{\bf IGR J17331-2406} & 263.306 & -24.156 & 1.8 &  0.6$\pm$0.1 &  0.8$\pm$0.1 & ? &  15.4 & 4191.0& S227B3 \\ 
4U 1730-335        & 263.351 & -33.386 & 0.5 &  5.9$\pm$0.1 &  2.6$\pm$0.1 & LMXB,G,RB,T &  81.1 & 4549.0& S100B1 \\ 
{\bf IGR J17348-2045} & 263.699 & -20.750 & 4.3 & $<$ 0.2 &  0.7$\pm$0.2 & ? &   5.8 & 3348.0& R230B2 \\ 
{\bf IGR J17354-3255} & 263.839 & -32.937 & 2.2 &  1.2$\pm$0.1 &  0.7$\pm$0.1 & ? &  12.5 & 4436.0& B5 \\ 
GRS 1734-292        & 264.377 & -29.137 & 0.6 &  5.5$\pm$0.1 &  4.5$\pm$0.1 & AGN,Sy1 &  67.0 & 4858.0& B5 \\ 
{\bf IGR J17379-3747} & 264.473 & -37.783 & 5.3 &  0.3$\pm$0.1 &  0.4$\pm$0.2 & ? &   4.6 & 3134.0& R164B1 \\ 
SLX 1735-269        & 264.572 & -26.993 & 0.4 & 10.3$\pm$0.1 &  8.6$\pm$0.1 & LMXB,B & 123.5 & 4662.0& B5 \\ 
4U 1735-444        & 264.741 & -44.452 & 0.3 & 29.9$\pm$0.2 &  0.9$\pm$0.2 & LMXB,B,A & 231.6 & 1411.0& B4 \\ 
XTE J17391-3021        & 264.800 & -30.349 & 1.2 &  1.5$\pm$0.1 &  0.8$\pm$0.1 & HMXB,SFXT,Be? &  24.0 & 5132.0& R106B1 \\ 
{\bf AX J1739.3-2923} & 264.875 & -29.370 & 4.9 &  0.3$\pm$0.1 &  0.5$\pm$0.1 & ? &   5.1 & 4858.0& B2 \\ 
XTE J1739-285        & 264.985 & -28.488 & 1.1 &  1.5$\pm$0.1 &  0.9$\pm$0.1 & LMXB,B,T &  26.3 & 4870.0& R362B2 \\ 
{\bf IGR J17404-3655} & 265.112 & -36.913 & 3.5 &  0.8$\pm$0.1 &  0.7$\pm$0.2 & ? &   7.3 & 3139.0& B5 \\ 
{\bf IGR J17407-2808} & 265.173 & -28.202 & 4.2 &  2.2$\pm$0.1 &  2.6$\pm$0.1 & ?,SFXT? &   6.0 & 4639.0& R243B1 \\ 
SLX 1737-282        & 265.179 & -28.291 & 1.3 &  3.7$\pm$0.1 &  3.8$\pm$0.1 & LMXB,B &  22.5 & 4699.0& S227B3 \\ 
2E 1739.1-1210        & 265.474 & -12.215 & 2.7 &  1.4$\pm$0.2 &  1.8$\pm$0.3 & AGN,Sy1 &   9.9 & 1274.0& B5 \\ 
{\bf IGR J17419-2802} & 265.486 & -28.034 & 2.0 &  0.3$\pm$0.1 &  0.3$\pm$0.1 & ?,T &  13.4 & 4482.0& R362B1 \\ 
{\bf IGR J17426-0258} & 265.645 & -2.963 & 4.3 & $<$ 0.5 & $<$ 0.9 & ? &   5.9 & 526.0& R429B1 \\ 
XTE J1743-363        & 265.747 & -36.377 & 1.1 &  3.2$\pm$0.1 &  2.5$\pm$0.1 & ? &  27.0 & 3213.0& B5 \\ 
1E 1740.7-2942        & 265.978 & -29.750 & 0.2 & 29.8$\pm$0.1 & 36.6$\pm$0.1 & LMXB,BHC & 486.3 & 5143.0& S100B3 \\ 
IGR J17445-2747        & 266.120 & -27.756 & 2.3 &  0.5$\pm$0.1 &  0.4$\pm$0.1 & ? &  11.9 & 4383.0& S163B3 \\ 
{\bf IGR J17448-3232} & 266.229 & -32.550 & 2.2 &  0.8$\pm$0.1 &  0.5$\pm$0.1 & ? &  12.4 & 4920.0& B4 \\ 
KS 1741-293        & 266.234 & -29.352 & 0.6 &  5.2$\pm$0.1 &  4.2$\pm$0.1 & LMXB,B,T &  67.2 & 5174.0& B5 \\ 
IGR J17456-2901\tablenotemark{p}        & 266.410 & -29.021 & 0.6 &  5.3$\pm$0.1 &  2.9$\pm$0.1 & ? &  65.2 & 5203.0& B5 \\ 
1A 1742-294        & 266.523 & -29.517 & 0.8 & 14.8$\pm$0.1 &  7.5$\pm$0.1 & LMXB,B &  39.2 & 5188.0& R411B1 \\ 
{\bf IGR J17461-2853\tablenotemark{q}} & 266.523 & -28.882 & 0.6 &  5.5$\pm$0.1 &  3.1$\pm$0.1 & mol cloud? &  65.5 & 5096.0& B5 \\ 
{\bf IGR J17461-2204} & 266.533 & -22.059 & 3.7 &  0.5$\pm$0.1 &  0.5$\pm$0.1 & ? &   6.8 & 4165.0& B4 \\ 
IGR J17464-3213        & 266.565 & -32.233 & 0.2 & 27.8$\pm$0.1 & 20.9$\pm$0.1 & LMXB,BHC,T & 658.9 & 4858.0& S047B1 \\ 
1E 1743.1-2843\tablenotemark{r}        & 266.580 & -28.735 & 0.6 &  5.5$\pm$0.1 &  2.1$\pm$0.1 & LMXB? &  65.7 & 4913.0& B4 \\ 
SAX J1747.0-2853        & 266.712 & -28.888 & 0.9 &  3.6$\pm$0.1 &  1.9$\pm$0.1 & LMXB,B,T &  38.0 & 5019.0& S411B1 \\ 
{\bf IGR J17472+0701} & 266.797 &  7.018 & 5.3 &  2.3$\pm$0.4 & $<$ 1.5 & ? &   4.6 & 192.0& B1 \\ 
{\bf IGR J17473-2721} & 266.819 & -27.348 & 4.7 & $<$ 0.2 &  0.5$\pm$0.1 & ?,T &   5.3 & 4691.0& S308B1 \\ 
IGR J17475-2822        & 266.830 & -28.400 & 1.0 &  2.5$\pm$0.1 &  2.1$\pm$0.1 & mol cloud? &  33.0 & 4964.0& B5 \\ 
SLX 1744-299        & 266.858 & -30.021 & 0.4 &  9.5$\pm$0.1 &  5.7$\pm$0.1 & LMXB,B & 114.3 & 5164.0& B5 \\ 
{\bf IGR J17476-2253} & 266.906 & -22.887 & 1.8 &  1.3$\pm$0.1 &  1.5$\pm$0.1 & AGN?,QSO? &  15.3 & 4424.0& B3 \\ 
GX 3+1        & 266.980 & -26.562 & 0.3 & 11.7$\pm$0.1 &  0.9$\pm$0.1 & LMXB,B,A & 186.5 & 4464.0& B4 \\ 
1A 1744-361        & 267.057 & -36.130 & 1.3 &  0.7$\pm$0.1 &  0.9$\pm$0.1 & LMXB,B,T &  23.1 & 3110.0& S181B3 \\ 
{\bf IGR J17487-3124} & 267.172 & -31.382 & 3.0 &  0.6$\pm$0.1 &  1.1$\pm$0.1 & ? &   8.8 & 5031.0& B5 \\ 
IGR J17488-3253        & 267.217 & -32.926 & 1.2 &  2.0$\pm$0.1 &  2.6$\pm$0.1 & AGN,Sy1 &  24.9 & 4700.0& B5 \\ 
4U 1745-203        & 267.222 & -20.368 & 1.4 &  0.6$\pm$0.1 &  0.9$\pm$0.2 & LMXB,G,T &  20.7 & 3496.0& R120B1 \\ 
{\bf AX J1749.1-2733} & 267.273 & -27.554 & 1.5 &  1.5$\pm$0.1 &  1.5$\pm$0.1 & HMXB?,SFXT? &  19.7 & 4030.0& R110B1 \\ 
{\bf SLX 1746-331} & 267.457 & -33.200 & 1.0 &  1.5$\pm$0.1 &  2.5$\pm$0.1 & LMXB,BHC,T &  31.4 & 4631.0& S100B3 \\ 
4U 1746-37        & 267.550 & -37.048 & 0.8 &  3.8$\pm$0.1 & $<$ 0.3 & LMXB,G,B,A  &  39.9 & 2867.0& B4 \\ 
{\bf IGR J17507-2647} & 267.677 & -26.792 & 2.6 &  0.9$\pm$0.1 &  0.9$\pm$0.1 & ? &  10.4 & 4563.0& B5 \\ 
{\bf IGR J17507-2856} & 267.681 & -28.941 & 2.2 &  0.5$\pm$0.1 &  0.3$\pm$0.1 & ?,T &  12.3 & 5153.0& S227B3 \\ 
GRS 1747-312        & 267.684 & -31.311 & 1.7 &  1.3$\pm$0.1 &  1.0$\pm$0.1 & LMXB,G,T &  16.2 & 4993.0& B4 \\ 
IGR J17513-2011        & 267.823 & -20.204 & 1.7 &  1.5$\pm$0.1 &  1.7$\pm$0.2 & AGN,Sy1.9 &  16.2 & 3628.0& B3 \\ 
{\bf IGR J17515-1533} & 267.882 & -15.544 & 4.8 &  0.3$\pm$0.1 & $<$ 0.4 & ? &   5.2 & 2029.0& R423B1 \\ 
{\bf SWIFT J1753.5-0127} & 268.371 & -1.456 & 0.7 &  5.2$\pm$0.2 &  6.9$\pm$0.4 & LMXB,BHC,T &  46.0 & 620.0& R365B1 \\ 
IGR J17544-2619        & 268.586 & -26.324 & 1.6 &  1.0$\pm$0.1 &  0.3$\pm$0.1 & HMXB,SFXT &  18.2 & 5009.0& R113B2 \\ 
{\bf IGR J17586-2129} & 269.657 & -21.327 & 3.3 &  0.6$\pm$0.1 & $<$ 0.3 & ? &   7.7 & 4028.0& B5 \\ 
IGR J17597-2201        & 269.935 & -22.026 & 0.6 &  6.2$\pm$0.1 &  5.4$\pm$0.1 & LMXB,B,D &  66.7 & 3832.0& B5 \\ 
GX 5-1        & 270.283 & -25.081 & 0.2 & 56.0$\pm$0.1 &  3.5$\pm$0.1 & LMXB,Z & 889.3 & 4417.0& B4 \\ 
GRS 1758-258        & 270.303 & -25.746 & 0.2 & 58.8$\pm$0.1 & 75.3$\pm$0.1 & LMXB,BHC & 812.7 & 4708.0& B3 \\ 
GX 9+1        & 270.389 & -20.531 & 0.3 & 18.3$\pm$0.1 &  0.4$\pm$0.2 & LMXB,A & 275.2 & 3445.0& B4 \\ 
SAX J1802.7-2017        & 270.661 & -20.304 & 0.7 &  6.0$\pm$0.1 &  1.8$\pm$0.2 & HMXB,XP,T &  53.3 & 3469.0& B1 \\ 
IGR J18027-1455        & 270.685 & -14.916 & 1.5 &  2.5$\pm$0.1 &  2.6$\pm$0.2 & AGN,Sy1 &  18.7 & 2024.0& B5 \\ 
IGR J18048-1455        & 271.197 & -14.966 & 3.4 &  1.0$\pm$0.1 &  0.5$\pm$0.2 & HMXB &   7.5 & 2046.0& B4 \\ 
XTE J1807-294        & 271.748 & -29.409 & 1.1 &  0.8$\pm$0.1 &  0.8$\pm$0.1 & LMXB,msecXP,T &  28.1 & 4446.0& R046B1 \\ 
SGR 1806-20        & 272.154 & -20.413 & 0.8 &  3.6$\pm$0.1 &  4.5$\pm$0.2 & SGR &  38.9 & 3450.0& B3 \\ 
{\bf PSR J1811-1926} & 272.827 & -19.417 & 2.9 &  0.8$\pm$0.1 &  1.0$\pm$0.2 & PSR,SNR,PWN? &   9.0 & 3531.0& B3 \\ 
{\bf IGR J18134-1636} & 273.350 & -16.598 & 3.8 &  0.7$\pm$0.1 &  1.0$\pm$0.2 & ? &   6.7 & 2643.0& B5 \\ 
IGR J18135-1751        & 273.395 & -17.871 & 2.2 &  1.1$\pm$0.1 &  1.7$\pm$0.2 & SNR,PWN? &  12.2 & 3009.0& B3 \\ 
GX 13+1        & 273.628 & -17.158 & 0.3 & 13.8$\pm$0.1 &  2.4$\pm$0.2 & LMXB,B,A & 151.3 & 2668.0& B4 \\ 
1M 1812-121        & 273.775 & -12.099 & 0.3 & 26.7$\pm$0.1 & 26.6$\pm$0.2 & LMXB,B & 198.3 & 1694.0& B5 \\ 
GX 17+2        & 274.007 & -14.036 & 0.2 & 71.0$\pm$0.1 &  3.8$\pm$0.2 & LMXB,B,Z & 671.8 & 1880.0& B4 \\ 
{\bf IGR J18173-2509} & 274.328 & -25.151 & 2.2 &  1.5$\pm$0.1 &  0.5$\pm$0.1 & ? &  12.3 & 3583.0& B5 \\ 
{\bf XTE J1817-330} & 274.431 & -33.025 & 0.3 &  6.9$\pm$0.1 &  4.2$\pm$0.2 & XB,BHC,T & 161.7 & 3637.0& R408B1 \\ 
SAX J1818.6-1703        & 274.657 & -17.045 & 0.9 &  1.6$\pm$0.1 &  1.3$\pm$0.2 & HMXB,SFXT &  34.3 & 2495.0& R110B1 \\ 
AX J1820.5-1434        & 275.126 & -14.568 & 1.1 &  2.8$\pm$0.1 &  1.8$\pm$0.2 & HMXB,XP,Be &  27.6 & 1869.0& S047B1 \\ 
IGR J18214-1318        & 275.335 & -13.330 & 2.3 &  1.7$\pm$0.1 &  1.7$\pm$0.2 & ?,T &  11.6 & 1718.0& B5 \\ 
4U 1820-303        & 275.917 & -30.362 & 0.2 & 35.6$\pm$0.1 &  2.0$\pm$0.2 & LMXB,G,B,A  & 442.4 & 3430.0& B4 \\ 
{\bf IGR J18244-5622} & 276.062 & -56.363 & 5.0 &  1.7$\pm$0.4 & $<$ 1.3 & AGN,Sy2 &   4.9 & 247.0& B5 \\ 
{\bf IGR J18249-3243} & 276.148 & -32.701 & 4.2 &  0.6$\pm$0.1 &  0.8$\pm$0.2 & AGN &   6.0 & 2887.0& B2 \\ 
{\bf IGR J18246-1425} & 276.161 & -14.418 & 4.4 &  1.1$\pm$0.1 &  0.6$\pm$0.2 & ? &   5.7 & 1890.0& R308B1 \\ 
4U 1822-000        & 276.351 & -0.013 & 1.4 &  2.0$\pm$0.2 & $<$ 0.5 & LMXB &  20.3 & 1433.0& B4 \\ 
IGR J18256-1035        & 276.437 & -10.563 & 3.7 &  1.0$\pm$0.1 & $<$ 0.5 & ? &   6.8 & 1711.0& B1 \\ 
3A 1822-371        & 276.449 & -37.108 & 0.3 & 34.1$\pm$0.1 &  3.9$\pm$0.2 & LMXB,D & 254.5 & 2477.0& B4 \\ 
IGR J18259-0706        & 276.495 & -7.136 & 3.2 &  1.0$\pm$0.1 &  0.9$\pm$0.2 & AGN? &   8.1 & 1570.0& B5 \\ 
{\bf RX J1826.2-1450} & 276.523 & -14.833 & 3.2 &  1.0$\pm$0.1 &  1.7$\pm$0.2 & HMXB,microQSO &   8.0 & 2014.0& B3 \\ 
Ginga 1826-24        & 277.371 & -23.793 & 0.2 & 86.8$\pm$0.1 & 69.1$\pm$0.2 & LMXB,B & 801.7 & 3252.0& B5 \\ 
{\bf AX J183039-1002} & 277.666 & -10.007 & 4.8 &  0.8$\pm$0.1 & $<$ 0.5 & ? &   5.2 & 1677.0& B1 \\ 
{\bf IGR J18308-1232} & 277.696 & -12.532 & 3.3 &  0.8$\pm$0.1 &  1.0$\pm$0.2 & ? &   7.7 & 1784.0& B5 \\ 
IGR J18325-0756        & 278.117 & -7.946 & 1.3 &  2.5$\pm$0.1 &  1.4$\pm$0.2 & ?,T &  21.8 & 1660.0& S047B1 \\ 
SNR 021.5-00.9        & 278.388 & -10.579 & 1.2 &  3.3$\pm$0.1 &  3.3$\pm$0.2 & SNR,PWN &  24.7 & 1692.0& B5 \\ 
PKS 1830-211        & 278.419 & -21.073 & 1.2 &  2.6$\pm$0.1 &  3.4$\pm$0.2 & AGN,QSO &  24.4 & 2611.0& B3 \\ 
{\bf 3C382} & 278.786 & 32.707 & 4.4 &  3.4$\pm$1.4 &  4.9$\pm$1.8 & AGN,Sy1 &   5.6 &  34.0& R022B1 \\ 
XB 1832-330        & 278.929 & -32.986 & 0.5 & 10.9$\pm$0.1 & 10.5$\pm$0.2 & LMXB,G,B,T &  92.1 & 2576.0& B5 \\ 
AX J1838.0-0655        & 279.509 & -6.916 & 1.5 &  1.9$\pm$0.1 &  3.0$\pm$0.2 & SNR,PWN? &  18.6 & 1667.0& B3 \\ 
ESO 103-35        & 279.563 & -65.450 & 3.9 &  5.3$\pm$0.9 &  4.7$\pm$1.5 & AGN,Sy2 &   6.5 &  44.0& B5 \\ 
Ser X-1        & 279.992 &  5.036 & 0.4 & 11.8$\pm$0.1 &  0.6$\pm$0.2 & LMXB,B & 121.0 & 1750.0& B4 \\ 
AX J1841.0-0535        & 280.243 & -5.580 & 2.9 &  1.1$\pm$0.1 &  0.7$\pm$0.2 & HMXB,XP,Be?,SFXT &   9.0 & 1702.0& R429B2 \\ 
Kes 73        & 280.338 & -4.948 & 1.3 &  2.2$\pm$0.1 &  4.2$\pm$0.2 & SNR,AXP &  23.4 & 1777.0& B3 \\ 
{\bf 3C 390.3} & 280.586 & 79.781 & 2.2 &  2.9$\pm$0.3 &  4.1$\pm$0.5 & AGN,Sy1 &  12.6 & 306.0& B5 \\ 
IGR J18450-0435        & 281.259 & -4.567 & 2.3 &  1.6$\pm$0.1 &  1.2$\pm$0.2 & HMXB,SFXT &  11.7 & 1784.0& B5 \\ 
Ginga 1843+009        & 281.404 &  0.865 & 0.6 &  5.0$\pm$0.1 &  3.7$\pm$0.2 & HMXB,XP,Be,T  &  55.5 & 1910.0& S308B3 \\ 
AX J1846.4-0258        & 281.596 & -2.983 & 1.8 &  1.7$\pm$0.1 &  2.4$\pm$0.2 & SNR,PWN,AXP &  15.6 & 1795.0& B3 \\ 
IGR J18483-0311        & 282.068 & -3.171 & 0.8 &  4.7$\pm$0.1 &  2.6$\pm$0.2 & ? &  42.7 & 1802.0& R429B1 \\ 
3A 1845-024        & 282.076 & -2.425 & 2.3 &  0.7$\pm$0.1 &  0.6$\pm$0.2 & HMXB,XP,Be,T &  11.4 & 1753.0& R231B1 \\ 
{\bf IGR J18485-0047} & 282.115 & -0.779 & 3.4 &  1.0$\pm$0.1 &  0.8$\pm$0.2 & ? &   7.6 & 1858.0& B5 \\ 
IGR J18490-0000        & 282.278 & -0.017 & 3.0 &  1.1$\pm$0.1 &  1.2$\pm$0.2 & ? &   8.6 & 1949.0& B5 \\ 
4U 1850-087        & 283.264 & -8.704 & 0.9 &  5.2$\pm$0.1 &  4.4$\pm$0.2 & LMXB,G,B &  38.3 & 1517.0& B5 \\ 
IGR J18539+0727        & 283.490 &  7.469 & 1.1 &  0.8$\pm$0.1 &  1.2$\pm$0.2 & XB,BHC,T &  27.8 & 2284.0& R062B1 \\ 
V1223 Sgr        & 283.755 & -31.154 & 0.8 &  7.4$\pm$0.2 &  3.0$\pm$0.3 & CV,IP &  41.3 & 1371.0& B1 \\ 
XTE J1855-026        & 283.878 & -2.608 & 0.5 & 12.2$\pm$0.1 &  7.2$\pm$0.2 & HMXB,XP,T &  91.0 & 1821.0& B5 \\ 
2E 1853.7+1534        & 283.970 & 15.618 & 2.4 &  1.5$\pm$0.2 &  1.4$\pm$0.2 & AGN,Sy1 &  11.2 & 1405.0& B5 \\ 
XTE J1858+034        & 284.678 &  3.439 & 0.3 & 13.9$\pm$0.1 &  1.7$\pm$0.2 & HMXB,XP,Be,T & 213.5 & 2419.0& R189B2 \\ 
{\bf HETE J1900.1-2455} & 285.035 & -24.924 & 0.6 &  6.6$\pm$0.2 &  5.6$\pm$0.3 & LMXB,msecXP,B &  56.1 & 1013.0& S411B3 \\ 
XTE J1901+014        & 285.417 &  1.448 & 1.1 &  3.0$\pm$0.1 &  2.8$\pm$0.2 & XB,BHC?,T &  27.4 & 2248.0& B5 \\ 
4U 1901+03        & 285.915 &  3.203 & 0.2 & 37.7$\pm$0.1 &  4.5$\pm$0.2 & HMXB,XP,T & 478.3 & 2483.0& S047B1 \\ 
{\bf IGR J19048-1240} & 286.205 & -12.661 & 4.2 &  0.6$\pm$0.2 & $<$ 0.6 & ? &   5.9 & 998.0& R365B1 \\ 
{\bf SGR 1900+14} & 286.851 &  9.311 & 2.1 &  1.3$\pm$0.1 &  1.1$\pm$0.2 & SGR &  13.3 & 2478.0& B5 \\ 
XTE J1908+094        & 287.225 &  9.384 & 0.9 &  1.7$\pm$0.1 &  1.9$\pm$0.2 & LMXB,BHC,T &  35.6 & 2471.0& S047B3 \\ 
4U 1907+097        & 287.411 &  9.830 & 0.3 & 18.6$\pm$0.1 &  1.9$\pm$0.2 & HMXB,XP,T & 179.3 & 2431.0& B4 \\ 
{\bf AXJ1910.7+0917} & 287.664 &  9.273 & 3.4 &  0.7$\pm$0.1 &  0.6$\pm$0.2 & ? &   7.5 & 2517.0& B5 \\ 
4U 1909+07        & 287.701 &  7.596 & 0.4 & 14.9$\pm$0.1 &  8.6$\pm$0.2 & HMXB,XP & 137.6 & 2560.0& B5 \\ 
Aql X-1        & 287.814 &  0.585 & 0.4 & 13.6$\pm$0.1 & 11.9$\pm$0.2 & LMXB,B,A,T & 133.7 & 1942.0& S308B3 \\ 
SS 433        & 287.956 &  4.983 & 0.5 & 10.4$\pm$0.1 &  5.2$\pm$0.2 & HMXB,SG,microQSO &  94.7 & 2560.0& B5 \\ 
IGR J19140+0951        & 288.516 &  9.878 & 0.5 &  8.9$\pm$0.1 &  5.6$\pm$0.2 & HMXB,SG &  81.5 & 2339.0& B5 \\ 
GRS 1915+105        & 288.799 & 10.944 & 0.2 & 296.8$\pm$0.1 & 123.4$\pm$0.2 & LMXB,BH,T & 2556.6 & 2342.0& B4 \\ 
4U 1916-053        & 289.701 & -5.238 & 0.6 &  9.9$\pm$0.2 &  5.4$\pm$0.3 & LMXB,B,D &  58.8 & 1147.0& B5 \\ 
{\bf SWIFT J1922.7-1716} & 290.633 & -17.305 & 3.8 &  1.0$\pm$0.3 & $<$ 0.9 & ? &   6.7 & 474.0& S308B3 \\ 
{\bf 1RXS J192450.8-291437} & 291.246 & -29.235 & 5.0 &  1.1$\pm$0.3 &  1.2$\pm$0.4 & AGN,BL Lac &   5.0 & 506.0& B3 \\ 
{\bf IGR J19267+1325} & 291.670 & 13.425 & 3.7 &  0.7$\pm$0.1 &  0.6$\pm$0.2 & ? &   6.8 & 1792.0& B5 \\ 
{\bf IGR J19378-0617} & 294.413 & -6.218 & 4.4 &  1.5$\pm$0.2 & $<$ 0.8 & AGN,Sy1 &   5.7 & 601.0& B5 \\ 
RX J1940.1-1025        & 295.058 & -10.428 & 3.0 &  2.7$\pm$0.3 &  2.0$\pm$0.5 & CV,P,asynch &   8.7 & 369.0& B1 \\ 
{\bf IGR J19405-3016} & 295.120 & -30.266 & 5.4 &  0.9$\pm$0.3 &  1.2$\pm$0.5 & AGN &   4.6 & 371.0& B1 \\ 
NGC 6814        & 295.657 & -10.320 & 2.4 &  3.2$\pm$0.3 &  3.7$\pm$0.6 & AGN,Sy1.5 &  11.4 & 329.0& B5 \\ 
{\bf IGR J19443+2117} & 296.069 & 21.287 & 4.9 &  1.4$\pm$0.3 &  0.9$\pm$0.4 & ? &   5.1 & 574.0& B3 \\ 
{\bf IGR J19473+4452} & 296.823 & 44.906 & 4.5 &  1.8$\pm$0.3 &  1.8$\pm$0.4 & AGN,Sy2 &   5.6 & 467.0& B3 \\ 
{\bf IGR J19487+5120} & 297.184 & 51.336 & 4.4 & $<$ 0.9 & $<$ 1.3 & ? &   5.7 & 240.0& R323B2 \\ 
KS 1947+300        & 297.395 & 30.209 & 0.8 & 13.5$\pm$0.3 &  9.7$\pm$0.4 & HMXB,XP,T &  43.3 & 502.0& B5 \\ 
{\bf 4U 1954+31} & 298.926 & 32.102 & 1.2 &  7.0$\pm$0.2 &  3.0$\pm$0.3 & LMXB,NS? &  25.0 & 677.0& B1 \\ 
Cyg X-1        & 299.590 & 35.199 & 0.2 & 763.7$\pm$0.2 & 876.7$\pm$0.3 & HMXB,BH,U & 4651.3 & 1142.0& B5 \\ 
Cyg A        & 299.868 & 40.749 & 1.4 &  4.8$\pm$0.2 &  4.8$\pm$0.3 & AGN,Sy2 &  20.1 & 789.0& B5 \\ 
{\bf SWIFT J2000.6+3210} & 300.085 & 32.177 & 2.9 &  2.0$\pm$0.2 &  1.8$\pm$0.3 & HMXB,Be &   8.9 & 734.0& B5 \\ 
{\bf ESO 399-20} & 301.691 & -34.559 & 4.9 &  1.7$\pm$0.4 &  1.4$\pm$0.7 & AGN,Sy1 &   5.1 & 241.0& B5 \\ 
{\bf IGR J20186+4043} & 304.690 & 40.703 & 3.3 &  1.3$\pm$0.2 &  1.3$\pm$0.3 & AGN? &   7.8 & 954.0& B5 \\ 
{\bf IGR J20286+2544} & 307.135 & 25.772 & 4.4 &  2.6$\pm$0.5 &  3.8$\pm$0.6 & AGN,Sy2 &   5.7 & 230.0& B3 \\ 
EXO 2030+375        & 308.059 & 37.638 & 0.3 & 32.4$\pm$0.2 & 16.3$\pm$0.3 & HMXB,XP,Be,T & 150.0 & 1009.0& B5 \\ 
Cyg X-3        & 308.108 & 40.956 & 0.2 & 196.5$\pm$0.2 & 78.3$\pm$0.3 & HMXB & 1096.2 & 1000.0& B4 \\ 
{\bf 4C 74.26} & 310.585 & 75.145 & 3.9 &  3.2$\pm$0.5 &  2.6$\pm$0.9 & AGN,QSO &   6.4 & 113.0& B5 \\ 
SAX J2103.5+4545        & 315.894 & 45.749 & 0.5 & 16.0$\pm$0.2 &  7.8$\pm$0.3 & HMXB,XP,Be,T &  87.2 & 988.0& B5 \\ 
{\bf S52116+81} & 318.492 & 82.072 & 3.5 &  2.5$\pm$0.4 &  2.3$\pm$0.7 & AGN,Sy1 &   7.3 & 199.0& B5 \\ 
{\bf IGR J21178+5139} & 319.436 & 51.663 & 4.0 &  1.2$\pm$0.2 &  1.4$\pm$0.3 & AGN? &   6.2 & 685.0& B3 \\ 
{\bf V2069 Cyg} & 320.906 & 42.278 & 4.9 &  0.9$\pm$0.2 & $<$ 0.5 & CV,IP &   5.1 & 901.0& B4 \\ 
IGR J21247+5058        & 321.172 & 50.972 & 1.0 &  6.0$\pm$0.2 &  6.7$\pm$0.3 & AGN,Sy1 &  31.7 & 768.0& B5 \\ 
{\bf IGR J21272+4241} & 321.792 & 42.692 & 5.5 &  0.9$\pm$0.2 & $<$ 0.5 & ? &   4.5 & 791.0& B1 \\ 
{\bf SWIFT J2127.4+5654} & 321.866 & 56.918 & 3.6 &  2.3$\pm$0.3 & $<$ 1.0 & AGN,NL Sy1 &   7.0 & 348.0& B1 \\ 
IGR J21335+5105        & 323.438 & 51.121 & 1.6 &  3.8$\pm$0.2 &  1.6$\pm$0.3 & CV,IP &  17.8 & 710.0& B5 \\ 
{\bf IGR J21347+4737\tablenotemark{s}} & 323.673 & 47.620 & 4.5 &  1.1$\pm$0.2 &  0.7$\pm$0.3 & ? &   5.6 & 845.0& B5 \\ 
{\bf RX J2135.9+4728\tablenotemark{t}} & 324.016 & 47.535 & 4.7 &  1.0$\pm$0.2 &  0.9$\pm$0.3 & AGN,Sy1 &   5.3 & 843.0& B1 \\ 
SS Cyg        & 325.691 & 43.583 & 1.6 &  3.8$\pm$0.2 &  1.6$\pm$0.3 & CV,DN &  17.8 & 745.0& B4 \\ 
Cyg X-2        & 326.169 & 38.320 & 0.3 & 28.5$\pm$0.2 &  2.9$\pm$0.3 & LMXB,B,Z & 202.1 & 730.0& B4 \\ 
{\bf NGC 7172} & 330.492 & -31.874 & 1.7 &  4.8$\pm$0.3 &  4.9$\pm$0.6 & AGN,Sy2 &  17.0 & 271.0& B5 \\ 
{\bf BL Lac} & 330.678 & 42.288 & 3.8 &  1.4$\pm$0.2 &  1.7$\pm$0.4 & AGN,BL Lac &   6.8 & 544.0& B5 \\ 
4U 2206+543        & 331.987 & 54.508 & 1.0 &  8.0$\pm$0.3 &  5.7$\pm$0.4 & HMXB,Be &  31.4 & 531.0& B5 \\ 
{\bf FO Aqr} & 334.478 & -8.317 & 5.1 &  3.6$\pm$0.7 & $<$ 2.8 & CV,IP &   4.8 &  84.0& B1 \\ 
{\bf IGR J22234-4116} & 335.850 & -41.262 & 5.3 & $<$ 0.9 &  3.2$\pm$0.9 & ? &   4.6 & 146.0& B2 \\ 
{\bf IGR J22292+6647} & 337.295 & 66.788 & 4.8 &  0.9$\pm$0.2 &  0.6$\pm$0.3 & AGN,RG &   5.1 & 988.0& B1 \\ 
{\bf NGC 7314} & 338.932 & -26.076 & 4.9 &  2.3$\pm$0.5 &  2.2$\pm$0.9 & AGN,Sy1.9 &   5.1 & 164.0& B5 \\ 
{\bf MR 2251-178} & 343.543 & -17.607 & 3.5 &  3.4$\pm$0.6 &  5.0$\pm$1.1 & AGN,Sy1 &   7.3 & 101.0& B5 \\ 
{\bf MCG-02-58-022} & 346.200 & -8.666 & 3.7 &  2.8$\pm$0.4 &  1.9$\pm$0.8 & AGN,Sy1.5 &   6.9 & 147.0& B1 \\ 
{\bf IGR J23130+8608} & 348.261 & 86.133 & 4.8 &  1.9$\pm$0.5 & $<$ 1.8 & ? &   5.2 & 134.0& B5 \\ 
{\bf NGC 7603} & 349.692 &  0.206 & 5.5 &  2.6$\pm$0.6 & $<$ 2.2 & AGN,Sy1.5 &   4.5 &  87.0& B1 \\ 
Cas A        & 350.848 & 58.815 & 0.9 &  4.1$\pm$0.1 &  2.4$\pm$0.2 & SNR &  34.2 & 1633.0& B4 \\ 
{\bf IGR J23308+7120} & 352.694 & 71.336 & 4.5 &  0.9$\pm$0.2 & $<$ 0.6 & AGN? &   5.5 & 982.0& B1 \\ 
{\bf IGR J23524+5842} & 358.111 & 58.700 & 4.0 &  0.7$\pm$0.1 &  0.8$\pm$0.2 & ? &   6.3 & 1789.0& B5 \\ 
\enddata 
\tablenotetext{a}{Names in bold face indicate new detections since second catalog} 
\tablenotetext{b}{Position errors expressed as radius of 90\% confidence circle in arcminutes} 
\tablenotetext{c}{Time-averaged flux expressed in units of mCrab; appropriate conversion factors are: (20-40 keV) 10 mCrab = $7.57 \times 10^{-11}$ erg cm$^{-2}$ s$^{-1} = 1.71 \times 10^{-3}$ ph cm$^{-2}$ s$^{-1}$; (40-100 keV) 10 mCrab = $9.42 \times 10^{-11}$ erg cm$^{-2}$ s$^{-1} = 9.67 \times 10^{-4}$ ph cm$^{-1}$ s$^{-1}$} 
\tablenotetext{d}{Source type classifications: A=Atoll source (neutron star); AGN=Active galactic nuclei; AXP=Anomalous X-ray pulsar; B=Burster (neutron star); Be=B-type emission-line star; BH=Black hole (confirmed mass evaluation); BHC=Black hole candidate; Cluster=Cluster of galaxies; CV=Cataclysmic variable; D=Dipping source; DN=Dwarf Nova; G=Globular Cluster X-ray source; GRB=Gamma-Ray Burst; HMXB=High-mass X-ray binary; IP=Intermediate Polar; LMXB=Low-mass X-ray binary; Mol Cloud=Molecular cloud; NS=Neutron Star; P=Polar; PSR=Radio pulsar; PWN=Pulsar wind nebula; QSO = Quasar; RG=Radio Galaxy; SFXT=Supergiant Fast X-ray Transient; SG=Supergiant; SGR=Soft gamma-ray repeater; SNR=Supernova remnant; Sy=Seyfert galaxy; Symb=Symbiotic star; T=Transient source; U=Ultrasoft source; XB=Galactic X-ray binary; XP=X-ray pulsar; Z=Z-type source (neutron star)} 
\tablenotetext{e}{Maximimum significance in a single map; see mapcode column to identify map with maximum significance} 
\tablenotetext{f}{Corrected on-source exposure (ksec)} 
\tablenotetext{g}{Map with maximum significance: B1=20-40 keV, B2=30-60 keV, B3=20-100 keV, B4=17-30 keV, B5=18-60 keV; a prefix of RXXX indicates detection in revolution XXX, SXXX indicates detection in revolution sequence beginning at revolution XXX; ST = Staring data} 
\tablenotetext{h}{Source type from Veron and Veron} 
\tablenotetext{i}{Possibly associated with BAL QSO, SDSS J03184-0015} 
\tablenotetext{j}{Possibly blended with ESO121-28, 4.8' away} 
\tablenotetext{k}{Eta Carinae or source therein} 
\tablenotetext{l}{Source type from CFA catalogue} 
\tablenotetext{m}{Significantly offset from original source position} 
\tablenotetext{n}{Blended with NGC 6221} 
\tablenotetext{o}{Blended with ESO 138-1} 
\tablenotetext{p}{Coincident with Sgr A*, but not unambiguously identified; in confused Galactic Center region} 
\tablenotetext{q}{Coincident with G0.13-0.13 molecular cloud; in confused Galactic Center region} 
\tablenotetext{r}{In confused Galactic Center region} 
\tablenotetext{s}{Blended with RX J2135.9+4728} 
\tablenotetext{t}{Blended with IGR J21347+4737} 
\end{deluxetable*} 

\end{document}